\documentclass[aps,prc,twocolumn,superscriptaddress]{revtex4-2}

\usepackage{amsmath} %overset
\usepackage{amsmath,amssymb,mathrsfs}
\usepackage[pdftex]{graphicx}
\usepackage[pdftex]{color}
\usepackage{bm}
\begin{document}

%\preprint{}

\title{Electroproduction of the $\Lambda/\Sigma^0$ hyperons at $Q^2\simeq0.5\ (\mathrm{GeV}/c)^2$ at forward angles}



%\thanks{}

\author{K.~Okuyama}
\affiliation{Department of Physics, Graduate School of Science, Tohoku University, Sendai, Miyagi 980-8578 Japan}
\author{K.~Itabashi}
\affiliation{WPI-QUP, KEK, Oho 1-1, Tsukuba, Ibaraki 305-0801,Japan}
\author{S.~Nagao}
\affiliation{Department of Physics, Graduate School of Science, The University of Tokyo, Hongo, Tokyo 113-0033 Japan}
\author{S.N.~Nakamura}
\affiliation{Department of Physics, Graduate School of Science, Tohoku University, Sendai, Miyagi 980-8578 Japan}
\affiliation{Department of Physics, Graduate School of Science, The University of Tokyo, Hongo, Tokyo 113-0033 Japan}
\author{K.N.~Suzuki}
\affiliation{Department of Physics, Graduate School of Science, Kyoto University, Kyoto, Kyoto 606-8502 Japan}
\author{T.~Gogami}
\affiliation{Department of Physics, Graduate School of Science, Kyoto University, Kyoto, Kyoto 606-8502 Japan}
\author{B.~Pandey}
\affiliation{Department of Physics, Hampton University, Hampton, Virginia 23668, USA}
\affiliation{Department of Physics \& Astronomy, Virginia Military Institute, Lexington, Virginia 24450, USA}
\author{L.~Tang}
\affiliation{Department of Physics, Hampton University, Hampton, Virginia 23668, USA}
\affiliation{Thomas Jefferson National Accelerator Facility, Newport News, Virginia 23606, USA}
%%added 2023/9/20
\author{P.~Byd\v{z}ovsk\'{y}}
\affiliation{Nuclear Physics Institute, ASCR, 25068 \v{R}e\v{z}/Prague, Czech Republic}
\author{D.~Skoupil}
\affiliation{Nuclear Physics Institute, ASCR, 25068 \v{R}e\v{z}/Prague, Czech Republic}
\author{T.~Mart}
\affiliation{Departemen Fisika, FMIPA, Universitas Indonesia, Depok 16424, Indonesia}
\author{D.~Abrams}
\affiliation{Department of Physics, University of Virginia, Charlottesville, Virginia 22904, USA}
\author{T.~Akiyama}
\affiliation{Department of Physics, Graduate School of Science, Tohoku University, Sendai, Miyagi 980-8578 Japan}
\author{D.~Androic}
\affiliation{Department of Physics \& Department of Applied Physics, University of Zagreb, HR-10000 Zagreb, Croatia}
\author{K.~Aniol}
\affiliation{Physics and Astronomy Department, California State University, Los Angeles, California 90032, USA}%8
\author{C.~Ayerbe~Gayoso}
\affiliation{Department of Physics, The College of William and Mary, Virginia 23185, USA}
\author{J.~Bane}
\affiliation{Department of Physics, University of Tennessee, Knoxville, Tennessee 37996, USA}
\author{S.~Barcus}
\affiliation{Department of Physics, The College of William and Mary, Virginia 23185, USA}
\author{J.~Barrow}
\affiliation{Department of Physics, University of Tennessee, Knoxville, Tennessee 37996, USA}
\author{V.~Bellini}
\affiliation{Istituto Nazionale di Fisica Nucleare, Sezione di Roma, 00185, Rome, Italy}
\author{H.~Bhatt}
\affiliation{Department of Physics, Mississippi State University, Mississippi State, Mississippi 39762, USA}
\author{D.~Bhetuwal}
\affiliation{Department of Physics, Mississippi State University, Mississippi State, Mississippi 39762, USA}
\author{D.~Biswas}
\affiliation{Department of Physics, Hampton University, Hampton, Virginia 23668, USA}
\author{A.~Camsonne}
\affiliation{Thomas Jefferson National Accelerator Facility, Newport News, Virginia 23606, USA}
\author{J.~Castellanos}
\affiliation{Department of Physics, Florida International University, Miami, Florida 33199, USA}
\author{J-P.~Chen}
\affiliation{Thomas Jefferson National Accelerator Facility, Newport News, Virginia 23606, USA}
\author{J.~Chen}
\affiliation{Department of Physics, The College of William and Mary, Virginia 23185, USA}
\author{S.~Covrig}
\affiliation{Thomas Jefferson National Accelerator Facility, Newport News, Virginia 23606, USA}
\author{D.~Chrisman}
\affiliation{Department of Physics and Astronomy, Michigan State University, East Lansing, Michigan 48824, USA}
\affiliation{National Superconducting Cyclotron Laboratory, Michigan State University, East Lansing, MI 48824, USA}
\author{R.~Cruz-Torres}
\affiliation{Department of Physics, Massachusetts Institute of Technology, Cambridge, Massachusetts 02139, USA}
\author{R.~Das}
\affiliation{Department of Physics, State University of New York, Stony Brook, New York 11794, USA}%17
\author{E.~Fuchey}
\affiliation{Department of Physics, University of Connecticut, Storrs, Connecticut 06269, USA}
\author{K.~Gnanvo}
\affiliation{Department of Physics, University of Virginia, Charlottesville, Virginia 22904, USA}
\author{F.~Garibaldi}
\affiliation{Istituto Nazionale di Fisica Nucleare, Sezione di Roma, 00185, Rome, Italy}%11
\affiliation{Instituto Superiore di Sanit\`{a}, 00161, Rome, Italy}%19

\author{T.~Gautam}
\affiliation{Department of Physics, Hampton University, Hampton, Virginia 23668, USA}
\author{J.~Gomez}
\affiliation{Thomas Jefferson National Accelerator Facility, Newport News, Virginia 23606, USA}
\author{P.~Gueye}
\affiliation{Department of Physics, Hampton University, Hampton, Virginia 23668, USA}
\affiliation{National Superconducting Cyclotron Laboratory, Michigan State University, East Lansing, MI 48824, USA}
\author{T.J.~Hague}
\affiliation{Department of Physics, Kent State University, Kent, Ohio 44242 USA}%20
\author{O.~Hansen}
\affiliation{Thomas Jefferson National Accelerator Facility, Newport News, Virginia 23606, USA}
\author{W.~Henry}
\affiliation{Thomas Jefferson National Accelerator Facility, Newport News, Virginia 23606, USA}
\author{F.~Hauenstein}
\affiliation{Department of Physics, Old Dominion University, Norfolk, Virginia 23529, USA}
\author{D.W.~Higinbotham}
\affiliation{Thomas Jefferson National Accelerator Facility, Newport News, Virginia 23606, USA}
\author{C.E.~Hyde}
\affiliation{Department of Physics, Old Dominion University, Norfolk, Virginia 23529, USA}
\author{M.~Kaneta}
\affiliation{Department of Physics, Graduate School of Science, Tohoku University, Sendai, Miyagi 980-8578 Japan}
\author{C.~Keppel}
\affiliation{Thomas Jefferson National Accelerator Facility, Newport News, Virginia 23606, USA}
\author{T.~Kutz}
\affiliation{Department of Physics, State University of New York, Stony Brook, New York 11794, USA}%17
\author{N.~Lashley-Colthirst}
\affiliation{Department of Physics, Hampton University, Hampton, Virginia 23668, USA}
\author{S.~Li}
\affiliation{Department of Physics, University of New Hampshire, Durham, New Hampshire 03824, USA}
\affiliation{Nuclear Science Division, Lawrence Berkeley National Laboratory, Berkeley, CA 94720, USA}
\author{H.~Liu}
\affiliation{Department of Physics, Columbia University, New York, New York 10027, USA}
\author{J.~Mammei}
\affiliation{Department of Physics and Astronomy, University of Manitoba, Winnipeg, Manitoba R3T 2N2, Canada}
\author{P.~Markowitz}
\affiliation{Department of Physics, Florida International University, Miami, Florida 33199, USA}
\author{R.E.~McClellan}
\affiliation{Thomas Jefferson National Accelerator Facility, Newport News, Virginia 23606, USA}
\author{F.~Meddi}
\affiliation{Istituto Nazionale di Fisica Nucleare, Sezione di Roma, 00185, Rome, Italy}
\affiliation{Sapienza University of Rome, I-00185, Rome, Italy}
\author{D.~Meekins}
\affiliation{Thomas Jefferson National Accelerator Facility, Newport News, Virginia 23606, USA}
\author{R.~Michaels}
\affiliation{Thomas Jefferson National Accelerator Facility, Newport News, Virginia 23606, USA}
\author{M.~Mihovilovi\v{c}}
\affiliation{Faculty of Mathematics and Physics, University of Ljubljana, 1000 Ljubljana, Slovenia}
\affiliation{Jo\v{z}ef Stefan Institute, Ljubljana, Slovenia}
\affiliation{Institut f\"{u}r Kernphysik, Johannes Gutenberg-Universit\"{a}t Mainz, DE-55128 Mainz, Germany}
\author{A.~Moyer}
\affiliation{Department of Physics, Christopher Newport University, Newport News, Virginia 23606, USA}
\author{D.~Nguyen}
\affiliation{Department of Physics, Massachusetts Institute of Technology, Cambridge, Massachusetts 02139, USA}
\affiliation{University of Education, Hue University, Hue City, Vietnam}
\author{M.~Nycz}
\affiliation{Department of Physics, Kent State University, Kent, Ohio 44242 USA}
\author{V.~Owen}
\affiliation{Department of Physics, The College of William and Mary, Virginia 23185, USA}
\author{C.~Palatchi}
\affiliation{Department of Physics, University of Virginia, Charlottesville, Virginia 22904, USA}
\author{S.~Park}
\affiliation{Department of Physics, State University of New York, Stony Brook, New York 11794, USA}%17
\author{T.~Petkovic}
\affiliation{Department of Physics \& Department of Applied Physics, University of Zagreb, HR-10000 Zagreb, Croatia}
\author{S.~Premathilake}
\affiliation{Department of Physics, University of Virginia, Charlottesville, Virginia 22904, USA}
\author{P.E.~Reimer}
\affiliation{Physics Division, Argonne National Laboratory, Lemont, Illinois 60439, USA}
\author{J.~Reinhold}
\affiliation{Department of Physics, Florida International University, Miami, Florida 33199, USA}
\author{S.~Riordan}
\affiliation{Physics Division, Argonne National Laboratory, Lemont, Illinois 60439, USA}
\author{V.~Rodriguez}
\affiliation{Divisi\'{o}n de Ciencias y Tecnologia, Universidad Ana G. M\'{e}ndez, Recinto de Cupey, San Juan 00926, Puerto Rico}
\author{C.~Samanta}
\affiliation{Department of Physics \& Astronomy, Virginia Military Institute, Lexington, Virginia 24450, USA}
\author{S.N.~Santiesteban}
\affiliation{Department of Physics, University of New Hampshire, Durham, New Hampshire 03824, USA}%22
\author{B.~Sawatzky}
\affiliation{Thomas Jefferson National Accelerator Facility, Newport News, Virginia 23606, USA}
\author{S.~\v{S}irca}
\affiliation{Faculty of Mathematics and Physics, University of Ljubljana, 1000 Ljubljana, Slovenia}
\affiliation{Jo\v{z}ef Stefan Institute, Ljubljana, Slovenia}
\author{K.~Slifer}
\affiliation{Department of Physics, University of New Hampshire, Durham, New Hampshire 03824, USA}%22
\author{T.~Su}
\affiliation{Department of Physics, Kent State University, Kent, Ohio 44242 USA}
\author{Y.~Tian}
\affiliation{Department of Physics, Syracuse University, New York, New York 10016, USA}
\author{Y.~Toyama}
\affiliation{Department of Physics, Graduate School of Science, Tohoku University, Sendai, Miyagi 980-8578 Japan}
\author{K.~Uehara}
\affiliation{Department of Physics, Graduate School of Science, Tohoku University, Sendai, Miyagi 980-8578 Japan}
\author{G.M.~Urciuoli}
\affiliation{Istituto Nazionale di Fisica Nucleare, Sezione di Roma, 00185, Rome, Italy}
\author{D.~Votaw}
\affiliation{Department of Physics and Astronomy, Michigan State University, East Lansing, Michigan 48824, USA}
\affiliation{National Superconducting Cyclotron Laboratory, Michigan State University, East Lansing, MI 48824, USA}
\author{J.~Williamson}
\affiliation{School of Physics \& Astronomy, University of Glasgow, Glasgow, G12 8QQ, Scotland, UK}
\author{B.~Wojtsekhowski}
\affiliation{Thomas Jefferson National Accelerator Facility, Newport News, Virginia 23606, USA}
\author{S.A.~Wood}
\affiliation{Thomas Jefferson National Accelerator Facility, Newport News, Virginia 23606, USA}
\author{B.~Yale}
\affiliation{Department of Physics, University of New Hampshire, Durham, New Hampshire 03824, USA}%22
\author{Z.~Ye}
\affiliation{Physics Division, Argonne National Laboratory, Lemont, Illinois 60439, USA}
\author{J.~Zhang}
\affiliation{Department of Physics, University of Virginia, Charlottesville, Virginia 22904, USA}
\author{X.~Zheng}
\affiliation{Department of Physics, University of Virginia, Charlottesville, Virginia 22904, USA}
\collaboration{JLab Hypernuclear Collaboration}
\noaffiliation


%\date{\today}

\begin{abstract}
In 2018, the E12-17-003 experiment was conducted at the Thomas Jefferson National Accelerator Facility (JLab) to explore the possible existence of an $nn\Lambda$ state in the reconstructed missing mass distribution from a tritium gas target \cite{suzuki_cross-section_2022,pandey_spectroscopic_2022}.
As part of this investigation, data were also collected using a gaseous hydrogen target, not only for a precise absolute mass scale calibration but also for the study of $\Lambda/\Sigma^0$ electroproduction. This dataset was acquired at $Q^2\simeq0.5$~$(\mathrm{GeV}/c)^2$, $W=2.14$~GeV, and $\theta_\mathrm{\gamma K}^\mathrm{c.m.}\simeq8$~deg. It covers forward angles where photoproduction data are scarce and a low-$Q^2$ region that is of interest for hypernuclear experiments. On the other hand, this kinematic region is at a slightly higher $Q^2$ than previous hypernuclear experiments, thus providing crucial information for understanding the $Q^2$ dependence of the differential cross sections for $\Lambda/\Sigma^0$ hyperon electroproduction. 
This paper reports on the $Q^2$ dependence of the differential cross section for the $e + p \to e' + K^+ + \Lambda/\Sigma^0$ reaction at $0.2-0.8$~$(\mathrm{GeV}/c)^2$, and provides comparisons with the currently available theoretical models.

\end{abstract}

%\keywords{}

\maketitle


%%%%%%%%%%%%%%
%%Introduction%%
\section{INTRODUCTION\label{sec: Introduction}}
Studying the production of hyperons and hypernuclei provides invaluable insights into baryon-baryon interactions with an extended ﬂavor, strangeness. Due to the short lifetime of the $\Lambda$ hyperon/hypernucleus ($\sim10^{-10}$~s), it cannot be observed as a stable state naturally. However, one can consider a $\Lambda$-hypernucleus as a stable object in view of the strong interaction. The development of accelerator facilities and detection techniques has made it possible to produce and study hyperons and hypernuclei in laboratories.

The study of hyperons and hypernuclei systems via the $(e,e'K^+)$ reaction at the Thomas Jefferson National Accelerator Facility (JLab) has been a cornerstone program since the mid 1990s.
Understanding the (un)polarized differential cross section for hyperon electroproduction is a fundamental observable to estimate the production yield of hypernuclei. However, experimental data on hyperon electroproduction under various kinematic settings are far from satisfactory. Therefore, predictions from theoretical models have become vital to supplement the data, in particular, at low $Q^2$ and forward angles.

Isobaric models, based on the effective Lagrangian using hadron degrees of freedom, play an important role: Kaon-Maid  (KM) \cite{mart_evidence_1999,lee_quasifree_2001}, Saclay-Lyon A (SLA) \cite{david_electromagnetic_1996, mizutani_off-shell_1998}, and other \cite{bydzovsky_models_2003,skoupil_photoproduction_2016,skoupil_photo-_2018,bydzovsky_kaon_2005,williams_hyperon_1992} models describe kaon-hyperon production with reasonable agreement compared to the existing experimental data.

In these isobaric models, background contribution from the $t$ channel and/or $u$ channel often becomes problematic when describing the kaon-hyperon channel. As a countermeasure, Regge-plus-resonance models, which introduces Regge pole exchange, have been recently applied to strangeness electroproduction with good results \cite{corthals_forward-angle_2006,decruz_bayesian_2012,bydzovsky_photoproduction_2019}.

\begin{figure}[tbp]
\includegraphics[width=90mm]{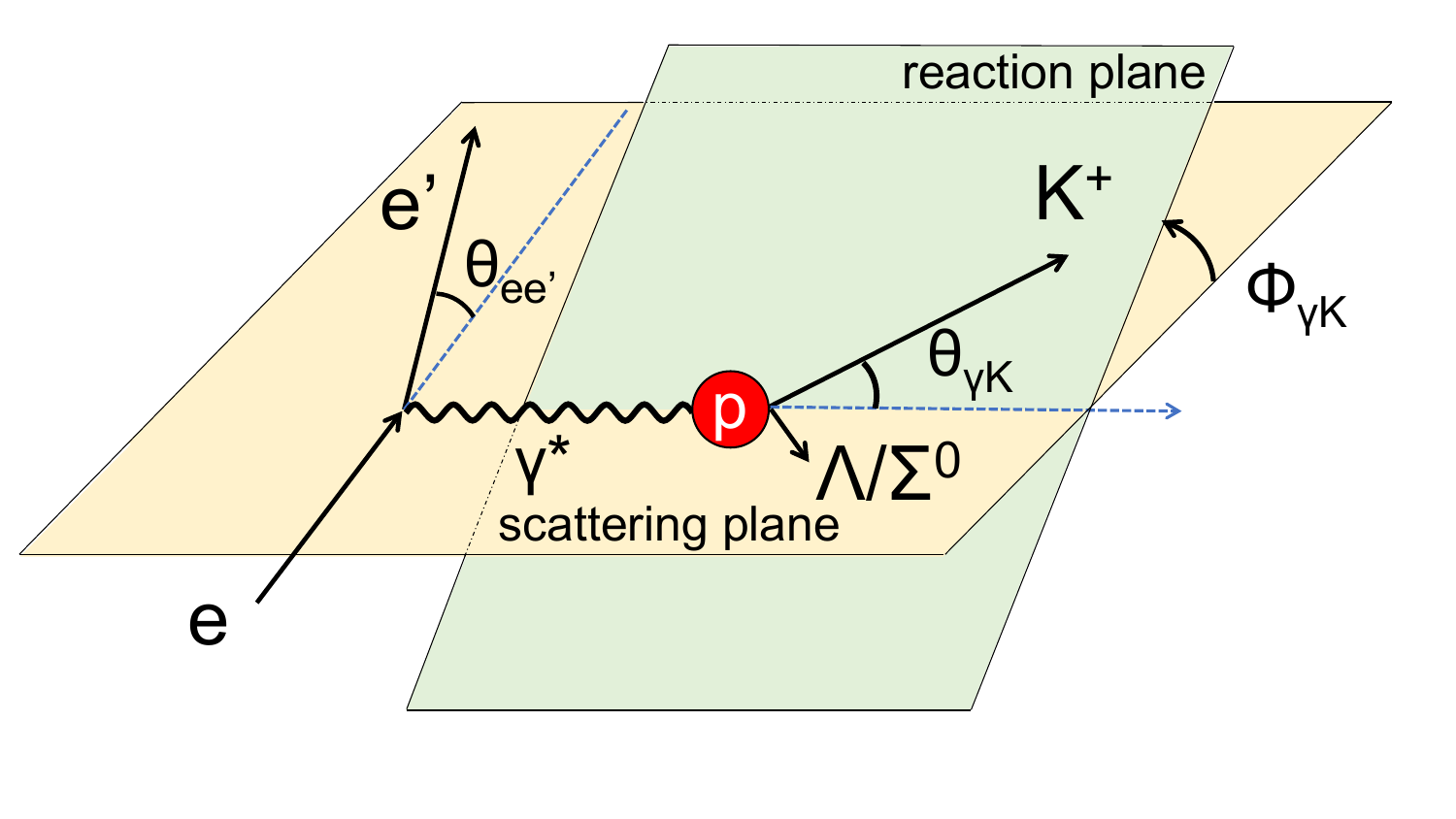}
\caption{Schematic drawing of the $p(e,e'K^+)\Lambda/\Sigma^0$ reaction. This shows the $\Lambda/\Sigma^0-K^+$ production under One-Photon-Exchange-Approximation. \label{fig: kinematics}}
\end{figure}

In the One-Photon-Exchange-Approximation, the electroproduction process $p(e,e'K^+)\Lambda/\Sigma^0$ can be related to the photoproduction one via a virtual photon $p(\gamma^*, K^+)\Lambda/\Sigma^0$ as shown in Fig. \ref{fig: kinematics}. This relation is given by \cite{amaldi_pion_1979},
\begin{align}
\frac{\mathrm{d}^3 \sigma}{\mathrm{d} E_{\mathrm{e}'} \mathrm{d} \Omega_{\mathrm{e}'}\mathrm{d}\Omega_{\mathrm{K}}^{\mathrm{c.m.}}}=\Gamma \frac{\mathrm{d}\sigma_{\gamma^*}}{\mathrm{d}\Omega^{\mathrm{c.m.}}_{\mathrm{K}}},
\end{align}
where $\Gamma$ is the so-called virtual photon flux. $\mathrm{d} \sigma_{\gamma^*}/\mathrm{d} \Omega_{\mathrm{K}}^\mathrm{c.m.}$ is regarded as a differential cross section for the kaon-hyperon production from a virtual photon. 
The four-momentum of a virtual photon is denoted as $q^\mu:=(\omega/c,\bm{q})=(E_\mathrm{e}/c-E_\mathrm{e'}/c,\bm{P}_\mathrm{e}-\bm{P}_\mathrm{e'}$).
The difference between photoproduction and electroproduction can be related using the four-momentum transfer, $Q^2:=-q^2=-\omega^2/c^2+|\bm{q}|^2$, i.e., $Q^2=0$ for photoproduction and $Q^2>0$ for electroproduction.

The differential cross section for virtual photoproduction can be decomposed into four terms when polarization observables are not measured in either the initial or the final state as in the present experiment:
\begin{align}
\frac{\mathrm{d}\sigma_{\gamma^*}}{\mathrm{d}\Omega_\mathrm{K}^\mathrm{c.m.}}&=\frac{\mathrm{d}\sigma_\mathrm{T}}{\mathrm{d}\Omega_\mathrm{K}^\mathrm{c.m.}}+\varepsilon\frac{\mathrm{d}\sigma_\mathrm{TT}}{\mathrm{d}\Omega_\mathrm{K}^\mathrm{c.m.}}\cos2\phi_\mathrm{\gamma \mathrm{K}}\nonumber\\&\quad+\varepsilon\frac{\mathrm{d}\sigma_\mathrm{L}}{\mathrm{d}\Omega_\mathrm{K}^\mathrm{c.m.}}+\sqrt{2\varepsilon\left(\varepsilon+1\right)}\frac{\mathrm{d}\sigma_\mathrm{LT}}{\mathrm{d}\Omega_\mathrm{K}^\mathrm{c.m.}}\cos\phi_{\gamma\mathrm{K}}.\label{eq: dcs_decomp}
\end{align}
Each term can be calculated using theoretical models and subsequently combined as in Eq. (\ref{eq: dcs_decomp}) to obtain $\mathrm{d} \sigma_{\gamma^*}/\mathrm{d} \Omega_{\mathrm{K}}^\mathrm{c.m.}$. 
In section \ref{sec: Discussion}, comparison of our experimental results to theoretical calculations will be discussed.

Hyperon electroproduction has been performed primarily at JLab, while hyperon photoproduction experiments have been performed by CLAS at JLab \cite{bradford_differential_2006,mccracken_differential_2010,mcnabb_hyperon_2004,dey_differential_2010}, SAPHIR at ELSA \cite{glander_measurement_2004}, LEPS at SPring-8 \cite{sumihama__2006}, and GRAAL at ESRF \cite{lleres_polarization_2007}. Experimentally, the photoproduction process has been well investigated providing abundant data for a wide range of angles to test theoretical models.
However, there are still large amounts of disagreements among the models due to the lack of data on photoproduction at forward and backward angles.
The electroproduction process has the advantage of acquiring data at forward angles, along the direction of virtual photons. 

The present paper is organized as follows. In section \ref{sec: Experiment}, an outline of our experiment is given. In section \ref{sec: Analysis}, event selection and derivation of the differential cross sections are explained. In section \ref{sec: Result}, the results for our $Q^2$ dependence are presented followed by a discussion in section \ref{sec: Discussion}. A conclusion is provided in section \ref{sec: Summary}.

%%%%%%%%%%%%%%%%%%%%%%%
%%Experimental Condition%%
%%%%%%%%%%%%%%%%%%%%%%%
\section{Experiment\label{sec: Experiment}}
In the present paper, the data using the gaseous hydrogen target in the  E12-17-003 experiment \cite{suzuki_cross-section_2022,pandey_spectroscopic_2022} ($Q^2 \simeq0.5\ (\mathrm{GeV}/ c)^2$, $ W = 2.14 $ GeV, and $\theta_\mathrm{\gamma K}^\mathrm{c.m.} \simeq8 $ deg) were analyzed, and the differential cross sections for the $p(e,e'K^+)\Lambda/\Sigma^0$ reaction were obtained.

\begin{figure}[tbp]
\includegraphics[width=0.48\textwidth]{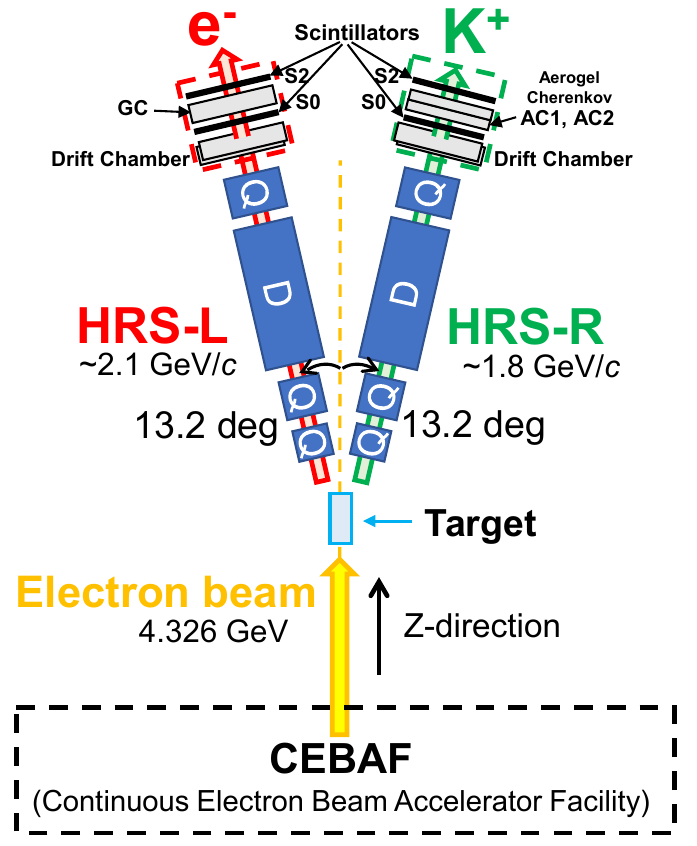}
\caption{Overall experimental setup (not to scale) of the E12-17-003 experiment at JLab Hall A. Quadrupole magnet (Q) and dipole magnet (D) are combined as QQDQ. $Z$-direction is defined as the electron beam direction. \label{fig: overall_apparatus}}
\end{figure}

The data were collected in Experimental Hall A using its two large magnetic spectrometers (HRS: High Resolution Spectrometer \cite{alcorn_basic_2004}) as shown in Fig. \ref{fig: overall_apparatus}. One of the HRSs was used to detect the scattered electrons (HRS-L), and the other was used to detect the produced kaons (HRS-R). Our experiment ran from October $31^\mathrm{th}$ to November $26^\mathrm{th}$ in 2018. 
We performed the experiment using an electron beam set at an energy of 4.326 GeV.
However, a more accurate beam energy of $\sim4.313$ GeV was measured periodically in front of the target, which was used in the present analysis. The kinematic settings of our experiment are summarized in Table \ref{tab: exp_info}.

We dedicated about $25\%$ of total beamtime into calibration runs using the hydrogen target (see Table \ref{tab: beam_info}).

\begin{table}[tbp]
\centering
\caption{Kinematic settings and experimental performance for the $p(e,e'K^+)\Lambda/\Sigma^0$ reaction using the hydrogen target in the E12-17-003 experiment}
\label{tab: exp_info}
\begin{tabular}{cc} \hline\hline
Electron Beam ($e$)& \\\hline
Beam Energy & 4.326 GeV\\
Energy Spread ($\Delta E/E$ in FWHM) & $\le1.0\times10^{-4}$ \\\hline
Scattered Electron ($e'$) & \\\hline
Central Momentum & $2.100$ GeV/$c$ \\
Momentum Acceptance & 4.5\%\\
Momentum Resolution ($\Delta p/p$ in FWHM) & $1.0\times10^{-4}$\\\hline 
Kaon ($K^+$)& \\\hline
Central Momentum & $1.823$ GeV/$c$ \\
Momentum Acceptance & 4.5\%\\
Momentum Resolution ($\Delta p/p$ in FWHM) & $1.0\times10^{-4}$\\\hline \hline
\end{tabular}
\end{table}

\begin{table}[tbp]
\centering
\caption{Electron beams irradiated to the targets}
\label{tab: beam_info}
\begin{tabular}{cccc} \hline\hline
Target & Beam Current & Beam Charge & $N_\mathrm{e}$ \\
 & [$\mu$A]  & [C] & \\ \hline 
Tritium &  22.5 & 16.9 &$1.1\times10^{20}$ \\ 
Hydrogen & 22.5 & 4.7  & $2.9\times10^{19}$ \\ \hline\hline
\end{tabular}
\end{table}

\section{Analysis\label{sec: Analysis}}
The missing mass technique was used to reconstruct the mass of the hyperon ($M_\mathrm{X}$) from the measured four-momenta of the electrons $(E_{\mathrm{e}'}/c,\bm{P}_{\mathrm{e}'})$ and kaons $(E_{\mathrm{K}}/c,\bm{P}_{\mathrm{K}})$. In the case of the proton target, the missing masses correspond to masses of produced hyperons:
\begin{align}
M_\mathrm{X}c^2&=\left[E_\mathrm{X}^2-\bm{P}_\mathrm{X}^2c^2\right]^{1/2}\nonumber\\
&=\left[\left(E_\mathrm{e}-E_{\mathrm{e}'}+M_{\mathrm{p}}c^2-E_{\mathrm{K}} \right)^2\right.\nonumber\\
&\quad\left.-\left(\bm{P}_\mathrm{e}-\bm{P}_{\mathrm{e}'}-\bm{P}_{\mathrm{K}}\right)^2c^2\right]^{1/2}.\label{eq: mm_def}
\end{align}

To obtain a background free missing mass spectrum, event selection procedures are necessary. As a first step, we selected the reaction point ($Z$-vertex) to reject events originating from the aluminum alloy of the gas cell end caps. A next step was to identify kaons among positively charged particles detected in the HRS-R. Kaon identification was successfully accomplished using the detector packages of the HRS-R, such as the two aerogel Cherenkov detectors with refractive indices of 1.015 and 1.055, respectively. Furthermore, time-of-flight measurements were performed with a simultaneous use of plastic scintillators behind both spectrometers (see Fig. \ref{fig: overall_apparatus}).

\subsection{Vertex selection\label{sec: vertex_sel}}
Electron scattering occurs along the electron beam, in the $Z$-direction inside the gas target region. The $Z$-vertex position was reconstructed using both HRSs independently.
Each spectrometer bends charged particles vertically along the momentum dispersive plane leading to distinct positions and angles at their respective focal planes. Meanwhile, the horizontal components at the focal planes have information of the $Z$-vertex because it is independent of the momentum dispersion.
The $Z$-vertex reconstruction was found to be $5$ mm for $1\sigma$ using one of the HRSs.

\begin{figure}[tbp]
\includegraphics[width=90mm]{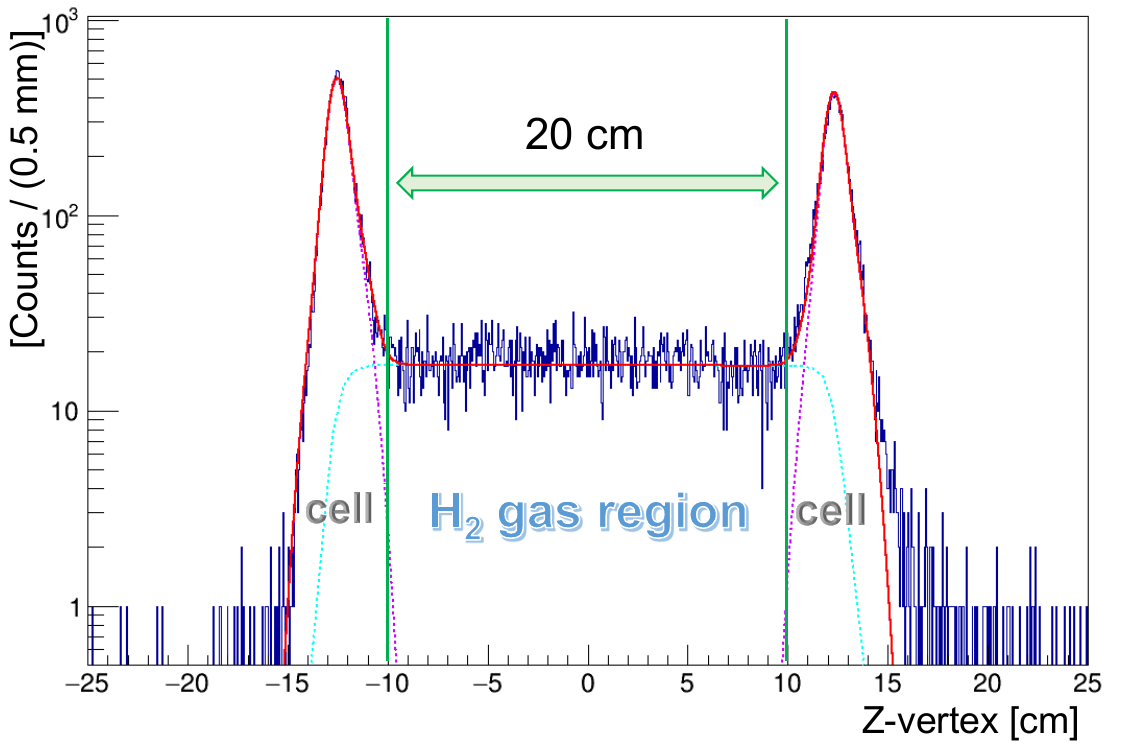}
\caption{The average $Z$-vertex distribution reconstructed by using the two HRSs. The green line shows the cut $|Z|<10\ \mathrm{cm}$ where events were selected. See text for details.
\label{fig: Al_fit_log}}
\end{figure}

The average $Z$-vertex distribution obtained from the HRSs and its fitting functions are shown in Fig. \ref{fig: Al_fit_log}: the two peaks at -12.5 cm (front) and +12.5 cm (rear) come from the target cell made of aluminum alloy \cite{santiesteban_density_2019}. Although the cell was designed to be 25 cm long, only events within $|Z|<10\ \mathrm{cm}$ were selected to avoid contamination. 

In Fig. \ref{fig: Al_fit_log}, the fitting functions consisted of double gaussians for the cell end caps and are shown in purple lines. A second-order polynomial function convoluted by a gaussian to include the gas region is shown as a cyan line. The aluminum contamination within the selected gas region was found to be about $0.3\%$.
The fitting worked well for $|Z|< 15$ cm; however, some events can be seen outside of the range. These events contribute to as much as 0.84$\%$ of total counts and were taken into account as a systematic error. Thus, the estimated Al contamination ratio within the selected region
was
\begin{align*}
(\mathrm{Al\ contami.\ ratio})=0.30\pm0.05(\mathrm{Stat.})^{+0.84}_{-0.00}(\mathrm{Syst.})\ [\%].
\end{align*}

\subsection{Kaon identification}
\begin{figure}
\includegraphics[width=90mm]{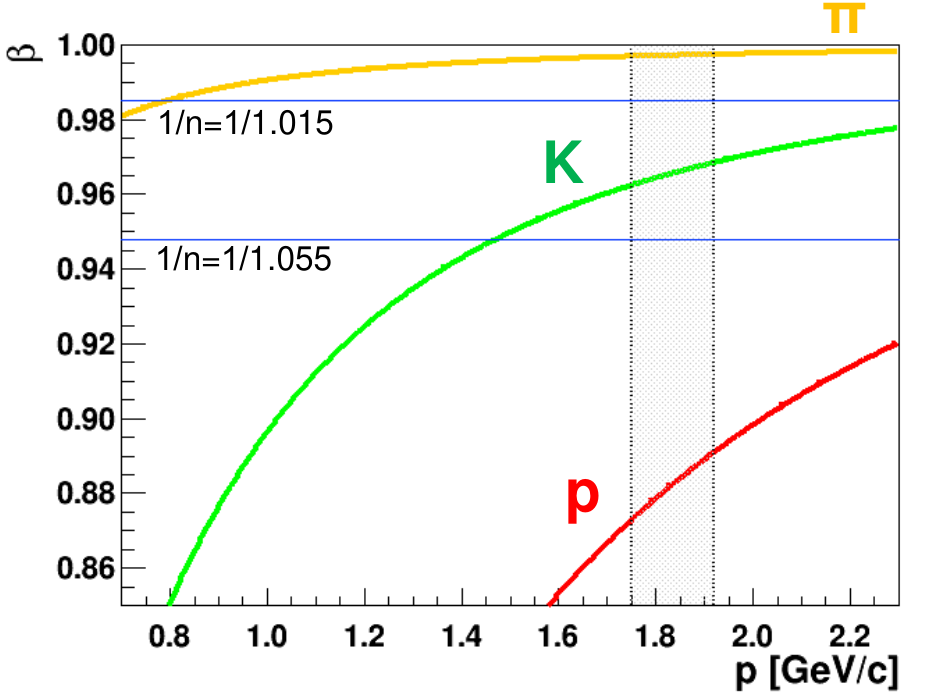}
\caption{$K/\pi/p$ discrimination using the Cherenkov detectors for this experiment (shaded band) \label{fig: ac_cherenkov_thre}}
\end{figure}

\begin{figure}[tb]
\includegraphics[width=90mm]{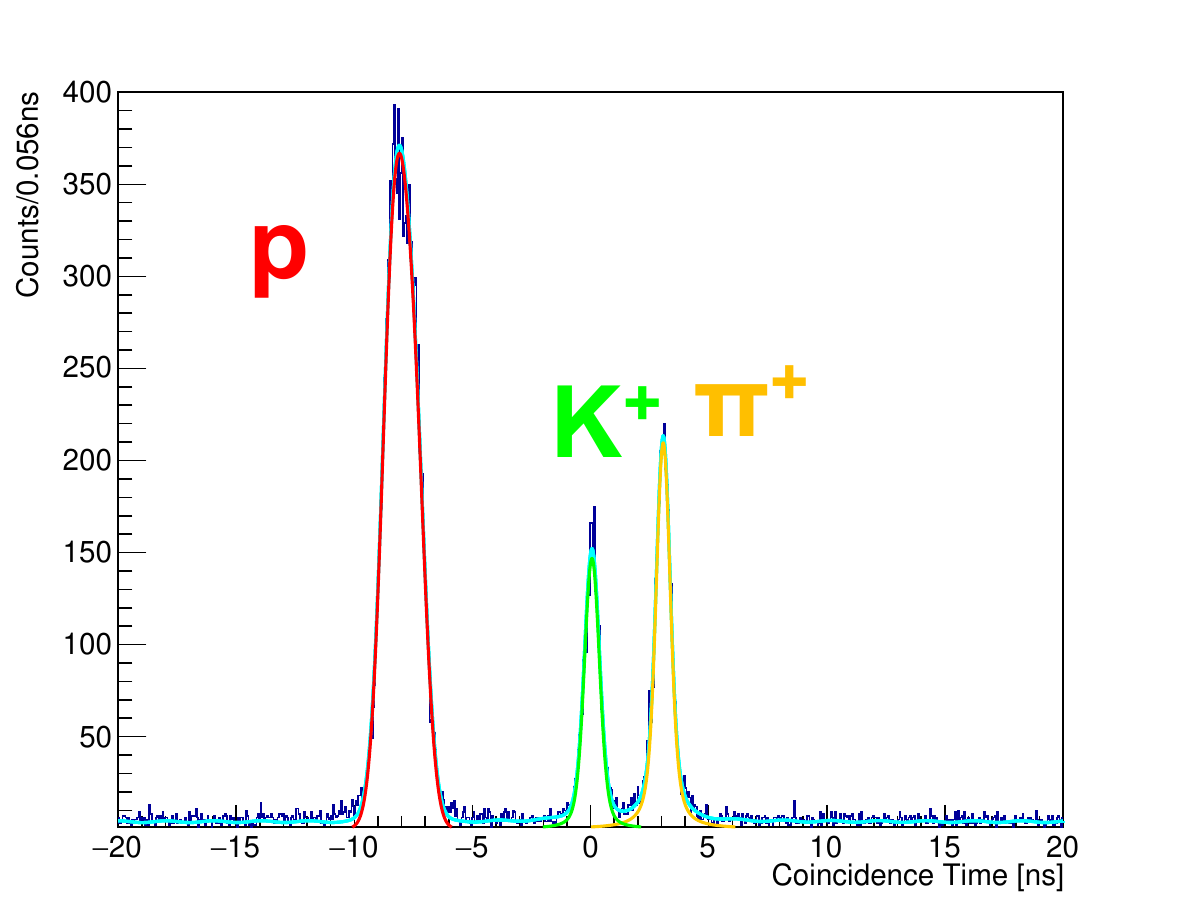}
\caption{Coincidence time distribution obtained from Eq. (\ref{eq: cointime2}). The kaon region is selected as $|t_\mathrm{Coin.}|<1\ \mathrm{ns}$. \label{fig: hcoin_fit}}
\end{figure}

Cherenkov detectors were used to separate $K^+$ from $\pi^+$ and $p$ based on their velocities in the HRS-R. The threshold for the Cherenkov light emission is $\beta_\mathrm{thres}=pc/\sqrt{m^2c^4+p^2c^2}>1/n$. In this experiment, $\sim1.8$-$\mathrm{GeV}/c$ particles passed through the aerogels located between the timing scintillators (see Fig. \ref{fig: overall_apparatus}). Figure \ref{fig: ac_cherenkov_thre} shows the relationship between momentum $p$ and the velocity  $\beta$ for $\pi^+/K^+$ and $K^+/p$ separation performed using the two aerogel detectors in the off-line analysis. However, large amounts of contamination from pions and protons still remain after applying the Cherenkov cut as indicated in Fig. \ref{fig: hcoin_fit}. In both spectrometers, the S2 plastic scintillators measure the time-of-flight from the target position:
\begin{align}
t(\mathrm{Target})&=t(\mathrm{S2})-\frac{\mathrm{Path\ Length}}{\beta c}\nonumber\\&=t(\mathrm{S2})-\frac{\sqrt{p^2c^2+m^2c^4}\times\mathrm{Path\ Length}}{pc^2}
\label{eq: cointime}
\end{align}
with a path length of about $27$ m. The coincidence time $t_\mathrm{Coin.}$ was defined as
\begin{align}
t_{\mathrm{Coin.}}:=t_{\mathrm{HRS\mathchar`-L}}(\mathrm{Target})-t_{\mathrm{HRS\mathchar`-R}}(\mathrm{Target}).
\label{eq: cointime2}
\end{align}
If we assume the mass $m$ is that of $K^+$ for HRS-R and that of $e$ for HRS-L, then $t_\mathrm{Coin.}=0$ for true $(e,e'K^+)$ events. The coincidence time distribution obtained from the hydrogen data is shown in Fig. \ref{fig: hcoin_fit}. Three distinct peaks corresponding to $\pi^+$, $K^+$, and $p$ are clearly seen. The underlying accidental background had a 2-ns bunch structure corresponding to the RF of the accelerator; however, it cannot be seen due to the strict cut of the aerogel Cherenkov detectors.
This accidental background will be discussed later in section \ref{sec: Missing mass spectrum}.
The cut condition for the coincidence time was chosen to be $|t_\mathrm{Coin.}|<1$ ns.

Figure \ref{fig: hcoin_fit} also shows the fitting functions used in the analysis taken to be a Voigt function for pions and kaons, and a double gaussian for protons. By definition, accidentals are periodic superpositions of these distributions functions. As a result of this fitting, the estimated pion contamination ratio within the selected region was
\begin{align*}
(\mathrm{\pi^+\ contami.\ ratio})=1.77^{+0.32}_{-0.28}(\mathrm{Stat.})^{+0.40}_{-0.04}(\mathrm{Syst.})\ [\%].
\end{align*}

\subsection{Missing mass spectrum\label{sec: Missing mass spectrum}}

\begin{figure}[tbp]
\includegraphics[width=90mm]{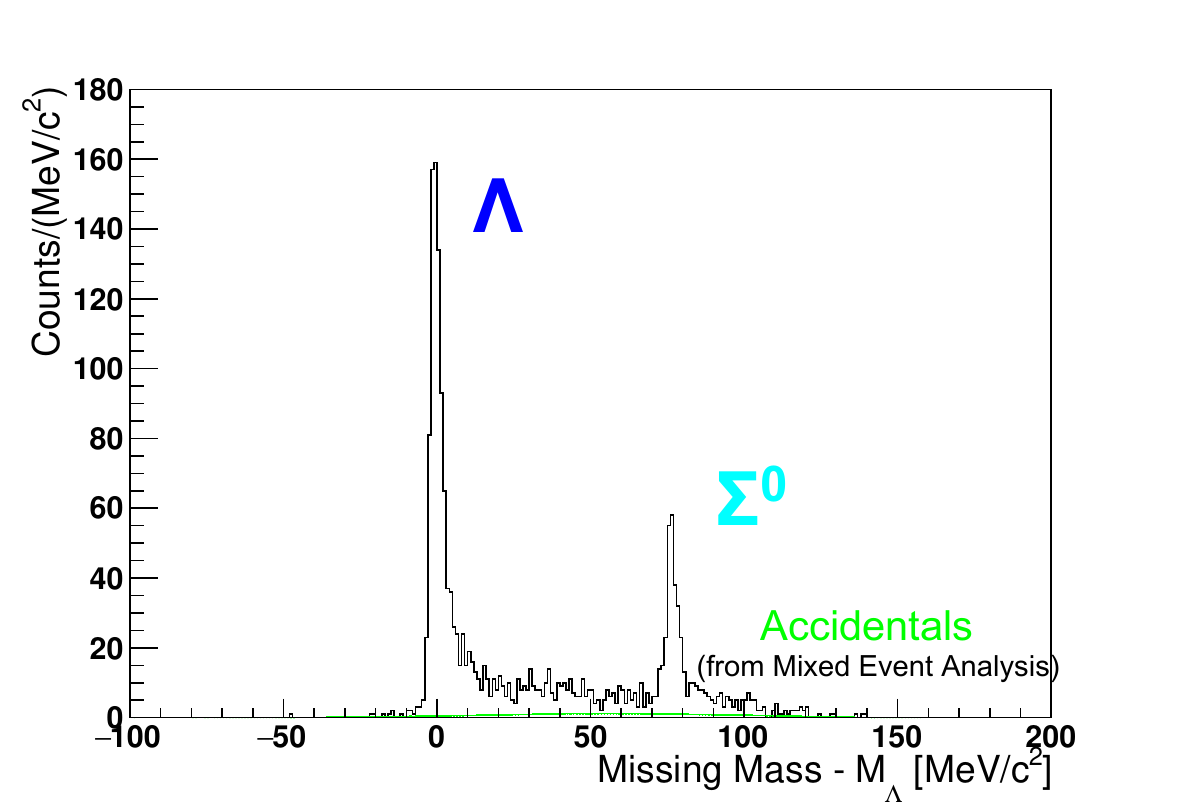}
\caption{Missing mass spectrum obtained from the $p(e,e'K^+)X$ reaction. \label{fig: hmm_rad_data}}
\end{figure}

The missing mass spectrum obtained from the hydrogen target is shown in Fig. \ref{fig: hmm_rad_data}. Kaon identification was successfully accomplished using the detector packages of HRS-R. As already shown in Fig. \ref{fig: hcoin_fit}, under such a high-rate continuous electron beam condition, background events due to accidental coincidences of scattered electrons in HRS-L and positively charged hadrons in HRS-R were unavoidable. 

However, the accidentals contribution can be deduced by making a distribution with artificially mixed events corresponding to random coincidences between the two HRS spectrometers. This analysis technique was applied and the result is shown in Fig. \ref{fig: hmm_rad_data} by the green line.

In Fig. \ref{fig: hmm_rad_data}, tail structures can be seen on the right side of both peaks corresponding to the radiative tails.
To derive the differential cross sections for the $\Lambda/\Sigma^0$ production, these radiative tails should be taken into account properly. Two techniques were used for this purpose; one was to fit the spectrum by using only the real data, and the other was to use a Monte-Carlo simulation (e.g. SIMC code \cite{noauthor_httpshallcwebjlaborgwikiindexphpsimc__nodate}). Both techniques are discussed in the following subsection.

\subsection{Radiative tail}
\begin{figure*}
\begin{minipage}[t]{0.48\textwidth}
%\begin{figure}[tb]
\includegraphics[width=90mm]{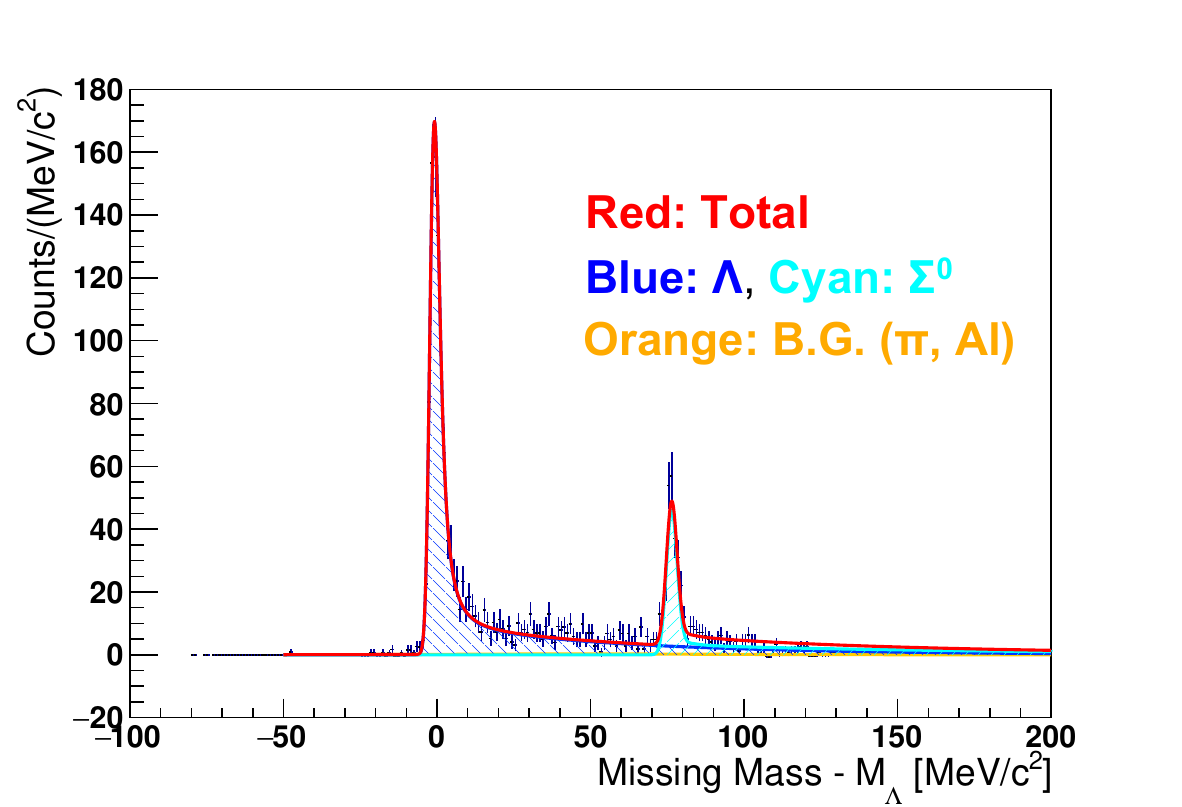}
\caption{Fitting result of the missing mass spectrum after accidentals subtraction. Blue and cyan lines are functions corresponding to the $\Lambda$ and $\Sigma^0$ production, respectively. The background and total fit function are shown in orange and red, respectively. \label{fig: hmm_rad_strict2}}
%\end{figure}
\end{minipage}
\hfill
\begin{minipage}[t]{0.48\textwidth}
%\begin{figure}[tb]
\includegraphics[width=90mm]{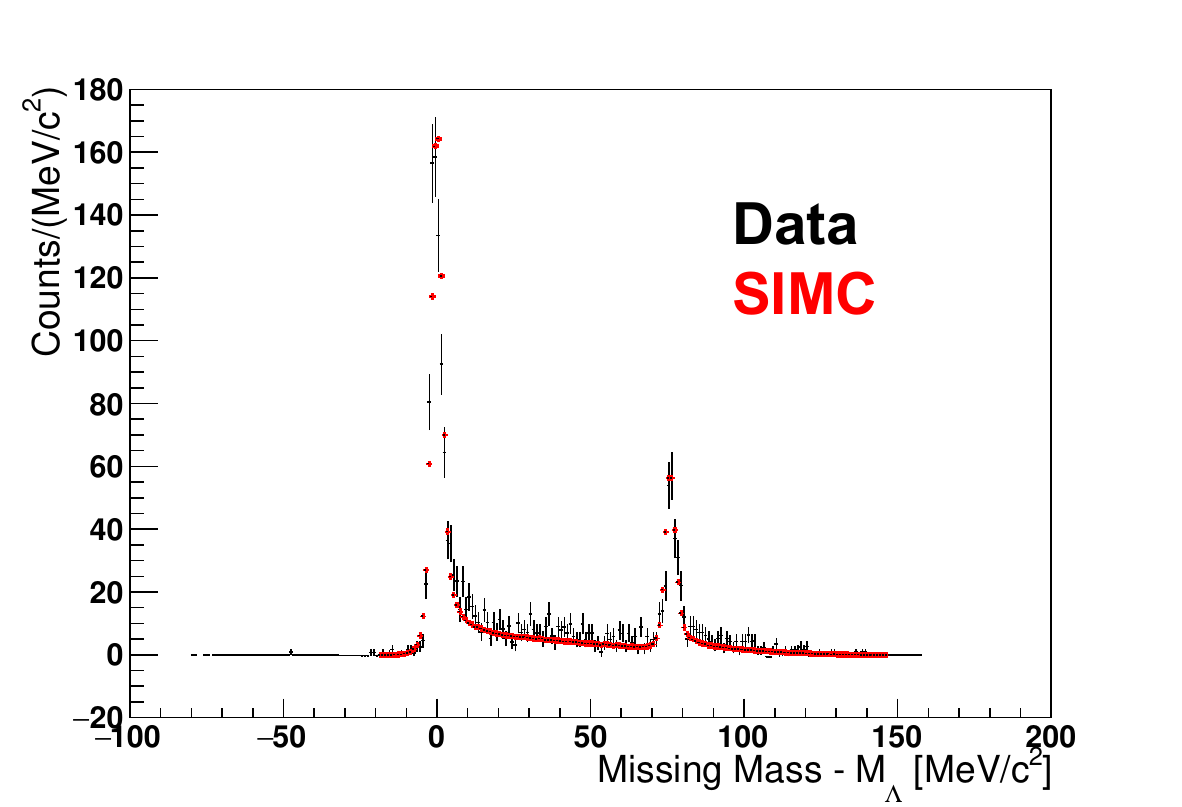}
\caption{Experimentally reconstructed missing mass spectrum (black points) compared to our SIMC Monte Carlo simulation (red points). \label{fig: mm_simc}}
%\end{figure}
\end{minipage}
\end{figure*}

The background subtracted missing mass spectrum is shown in Fig. \ref{fig: hmm_rad_strict2}. Other sources of background include events originating from the target cell windows in the $Z$-vertex cut (Fig. \ref{fig: Al_fit_log}), and pion contamination in the coincidence time cut (Fig. \ref{fig: hcoin_fit}). Proton contamination in the coincidence time cut was found to be negligible. The sum of those contributions was estimated to be $\sim2\%$.
Radiative tails can be seen on the right sides of the $\Lambda$ and $\Sigma^0$ peaks coming from both incoming and outgoing electrons (dominant contribution).

The numbers of $\Lambda$ and $\Sigma^0$ events included in the radiative tails can be obtained from the fit of the tails for both $\Lambda$ (blue) and $\Sigma^0$ (cyan) as shown in Fig. \ref{fig: hmm_rad_strict2}, respectively.

Each fit function consisted of a sum of a Landau distribution $h(x)$ and exponential function $f(x)$ convoluted by a Gaussian $g(x)$, namely, 
\begin{align}
((f+h)*g)(x)=(f*g)(x)+(h*g)(x)\label{eq: rad_fit_func_def}\\
\begin{cases}
f(x)&:\  \mathrm{exponential\ function}\nonumber\\
g(x)&:\  \mathrm{gaussian}\nonumber\\
h(x)&:\  \mathrm{Landau\ distribution}\nonumber
\end{cases}
\end{align}
Background contributions from the target cell and pion contamination are shown as the orange line originating from a fit with double Voigt functions. Scaling factors were determined based on the contamination ratios and the total fitting function is shown in a red line. The fitting result reproduced the data well.

Radiative tails are dominated by events from the incoming and outgoing electrons and were estimated using the in-house SIMC Monte Carlo simulation tool \cite{noauthor_httpshallcwebjlaborgwikiindexphpsimc__nodate}.
The full target geometry including its aluminum end caps as described in section \ref{sec: vertex_sel} was included in the simulation. Radiation effects from both internal \cite{vanderhaeghen_qed_2000} and external \cite{tsai_pair_1974} contributions were taken into account. Because of the limited solid angle of the HRS ($\sim5.5$~msr), particles were generated uniformly assuming a negligible angular dependence across the spectrometer acceptance. The results of the simulation including radiative effects are shown in Fig. \ref{fig: mm_simc}. 

The fitting result of Fig. \ref{fig: mm_simc} using a function that constructs the tail component only by radiative effects based on the simulation gives a smaller number of detected hyperons than that of Fig. \ref{fig: hmm_rad_strict2}, which reproduces the real data well.
This is because the data contain possible other unknown background. In the present analysis, the number of detected hyperons obtained from the fit of Fig. \ref{fig: hmm_rad_strict2} was chosen as the most faithful value based on the experimental data. Additionally, the variations of the fitting results by changing the fit conditions and the integral ranges, as well as those reflecting the distribution reproduced by SIMC, were taken into account as systematic errors. The upper error comes from the largest value among these fitting variations, and the lower error comes from the smallest value among them.

\subsection{Derivation of the differential cross section}
The extracted differential cross section was calculated from
\begin{align}
&\overline{\left( \frac{\mathrm{d}\sigma_{\gamma^*p\to K^+\Lambda(\Sigma^0)}}{\mathrm{d}\Omega_{K^+}} \right)}_{\mathrm{HRS\mathchar`-R},i}\nonumber\\
&=\frac{1}{N_{\mathrm{T}}}\cdot\frac{1}{N_{\gamma^*}}\cdot\frac{1}{\bar{\varepsilon}}\cdot\sum_{i=1}^{N_{\Lambda(\Sigma^0)}}\frac{1}{\varepsilon^\mathrm{DAQ}_i\cdot\varepsilon^{\mathrm{Decay}}_i\cdot\Delta\Omega_{\mathrm{HRS\mathchar`-R},i}}\label{eq: DCS_real}
\end{align}
where
\begin{align*}
N_{\Lambda(\Sigma^0)}&:\ \mathrm{Number\ of}\ \Lambda(\Sigma^0)\ \mathrm{events}\\
N_{\mathrm{T}}&:\ \mathrm{Number\ of\ proton\ targets\ } [\mathrm{b^{-1}}]\\
N_{\gamma^*}&:\ \mathrm{Number\ of\ virtual\ photons}\\
\bar{\varepsilon}&:\ \mathrm{Average\ event\ cut\ efficiency}\\
\varepsilon^{\mathrm{DAQ}}_i&:\ \mathrm{DAQ\ efficiency\ when\ taking}\ i\mathrm{\mathchar`-th\ event}\\
\varepsilon^{\mathrm{Decay}}_i&:\ \mathrm{Survival\ ratio\ of\ }K^+ \\ 
\Delta\Omega_{\mathrm{HRS\mathchar`-R},i}&:\ \mathrm{Solid\ angle\ with\ HRS\mathchar`-R\ for}\ i\mathrm{\mathchar`-th\ event\ }[\mathrm{sr}]
\end{align*}
Since the target thickness of the gaseous hydrogen was $70.8$ mg/cm$^2$, the number of the target is $N_\mathrm{T}=0.0375$ b$^{-1}$.
The efficiencies, survival ratio ($\varepsilon^\mathrm{Decay}\sim14\%$), and solid angle ($\Delta\Omega_{\mathrm{HRS\mathchar`-R}}\sim5.5$ msr) should be considered on an event by event basis; however, the cut efficiencies can be replaced by the average value $\bar{\varepsilon}$ obtained from the data. All components of the efficiency and the average value are summarized in Table \ref{tab: eff_info}. The kaon survival ratio $\varepsilon^\mathrm{Decay}$ and solid angle $\Delta\Omega_{\mathrm{HRS\mathchar`-R}}$ had a momentum dependence. Thus, these components were explicitly separated from $\bar{\varepsilon}$ and evaluated on an event by event basis.
The data acquisition efficiency $\varepsilon^\mathrm{DAQ}\sim96\%$ was evaluated on a run by run basis.
Additionally, Table \ref{tab: dcs_uncertainty} summarizes the estimated systematic errors.

%\begin{figure}[tbp]
%\includegraphics[width=70mm]{figures/acceptance2d_LHRS.pdf}
%\caption{$2$-dimensional plot of the solid angle (HRS-L) \label{fig: acceptance2d_LHRS}}
%\end{figure}
%\begin{figure}[tbp]
%\includegraphics[width=70mm]{figures/acceptance2d_RHRS.pdf}
%\caption{$2$-dimensional plot of the solid angle (HRS-R) \label{fig: acceptance2d_RHRS}}
%\end{figure}

\begin{table*}[tbp]%Wide table in two-column
\centering
\caption{Efficiencies used for the cross section estimation}
\label{tab: eff_info}
\begin{tabular}{lccl} \hline\hline
Item & $\bar{\varepsilon}(\Lambda)$ [$\%$] & $\bar{\varepsilon}(\Sigma^0)$ [$\%$] & Description \\ \hline
$\varepsilon^\mathrm{Z}$ & 82.5 & 76.2 & $Z$-vertex cut for hydrogen target selection\\
$\varepsilon^{\mathrm{AC}}$ & 60.2 & 59.1 & Aerogel Cherenkov cut for kaon identification\\
$\varepsilon^{\mathrm{CT}}$ & 98.8 & 97.0 & Coincidence Time cut for kaon identification\\
$\varepsilon^{\mathrm{Single}}$  & $97.0$ & $97.0$ & Percentage of single-hit events in HRS-L (excluding multi-hit)  \\ 
$\varepsilon^{\mathrm{FP}}$ & $96.2$ & $96.2$ & Focal Plane cut for removing unphysical events\\
$\varepsilon^{\mathrm{Track}}$ & $98.1$ & $98.1$ & Percentage of successful track reconstructions  \\ 
$\varepsilon^{\mathrm{\chi^2}}$  & $>99.9$ & $>99.9$ & $\chi^2$ cut for the reconstructed tracks  \\ \hline
$\varepsilon^{\mathrm{Total}}$  & $44.9$ & $40.0$ & Total efficiencies of the above \\ \hline\hline
\end{tabular}
\end{table*}

\begin{table}[tbp]
\centering
\caption{Summary of the estimated systematic errors}
\label{tab: dcs_uncertainty}
\begin{tabular}{lcccc} \hline\hline
 & \multicolumn{2}{c}{$\Lambda$} & \multicolumn{2}{c}{$\Sigma^0$}  \\
Item&Lower & Upper & Lower & Upper  \\ \hline
B.G. from Al Cell  & 0.89\% & 0.05\% & 0.89\% & 0.05\% \\
Pion Contamination & 0.72\% & 0.32\% & 0.72\% & 0.32\% \\
Radiative tail & 9.43\% & 4.30\% & 25.2\% & 47.6\% \\
Kaon Survival Ratio & 4.13\% & 0.76\% & 4.13\% & 0.76\% \\
Number of Target Centers & 0.83\% & 0.83\% & 0.83\% & 0.83\% \\
Number of Beam Particles & 1.00\% & 1.00\% & 1.00\% & 1.00\% \\
Number of Virtual Photons & 1.40\% & 2.30\% & 1.76\% & 1.11\% \\
Mixed Event Analysis  & 0.20\% & 0.20\% & 0.30\% & 0.30\% \\\hline
All & 10.53\% & 5.11\% & 25.7\% & 47.6\% \\\hline\hline
\end{tabular}
\end{table}

\section{Result\label{sec: Result}}
\begin{figure}[tbp]
\includegraphics[width=90mm]{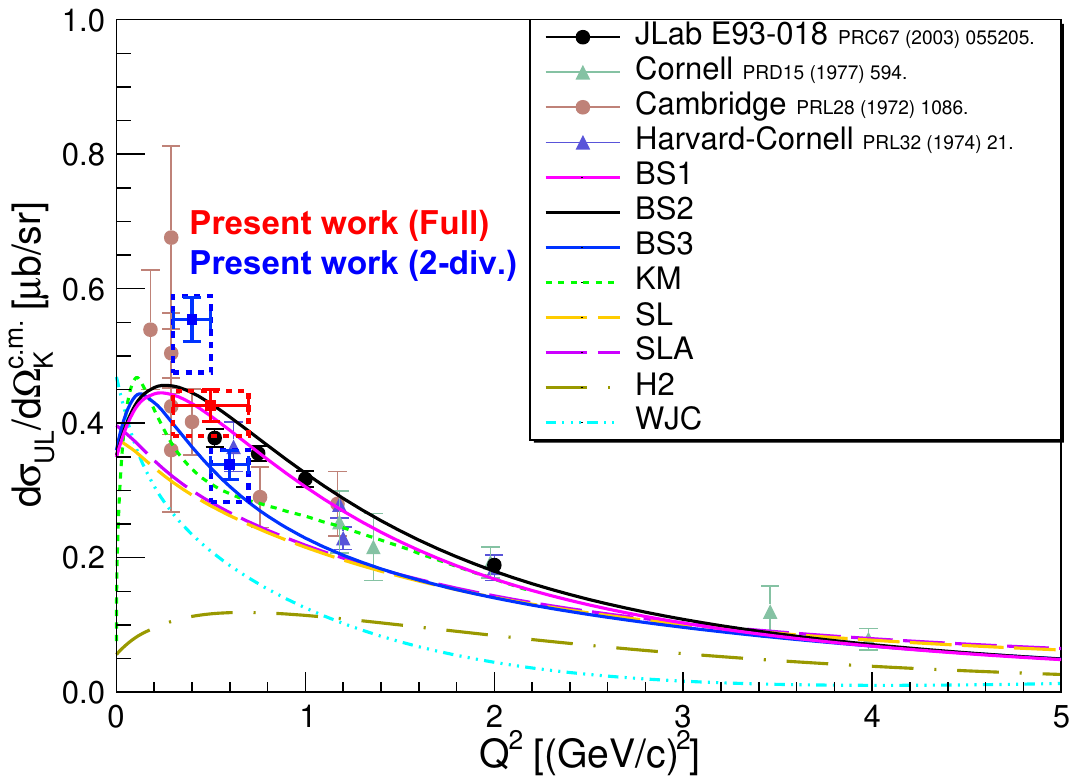}
\caption{$Q^2$-dependence of the differential cross section for the $p(\gamma^*,K^+)\Lambda$ reaction. \label{fig: CS_Q2depL2}}
\includegraphics[width=90mm]{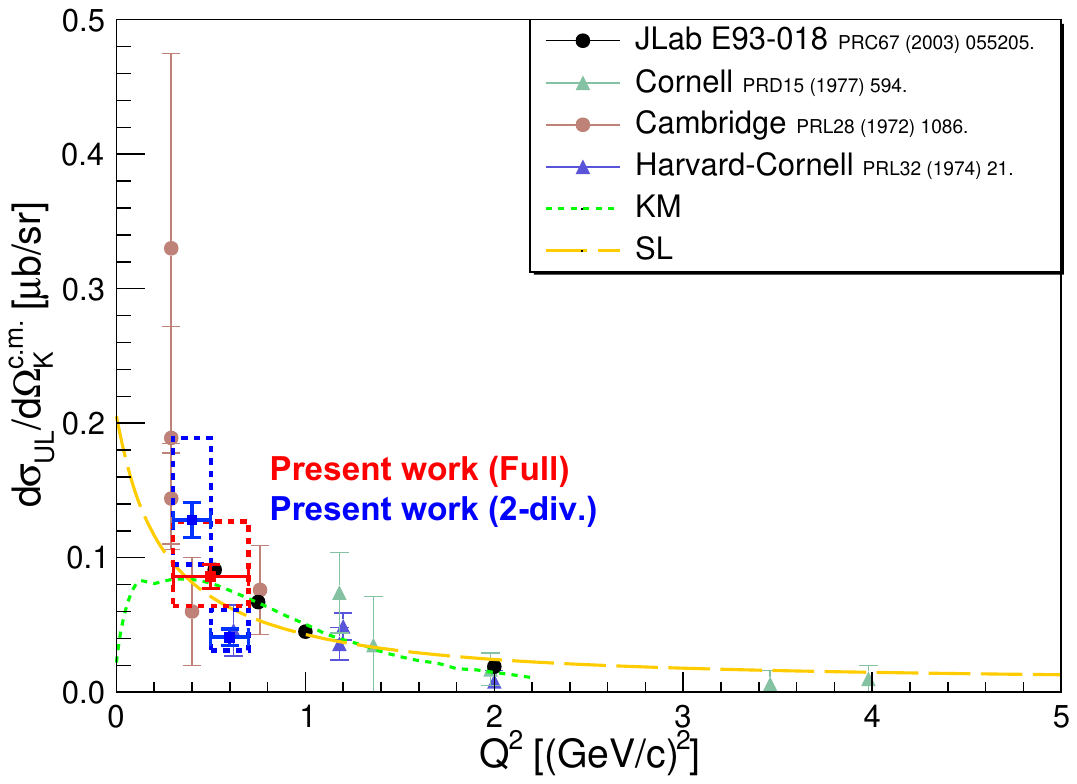}
\caption{$Q^2$-dependence of the differential cross section for the $p(\gamma^*,K^+)\Sigma^0$ reaction. \label{fig: CS_Q2depS2}}
\end{figure}

The measured differential cross sections for $\Lambda$ and $\Sigma^0$ are summarized in Table \ref{tab: dcs_results} including an analysis of our data using two regions: $Q^2<0.5\ (\mathrm{GeV}/c)^2$ and $Q^2\ge 0.5\ (\mathrm{GeV}/c)^2$ shown to provide a $Q^2$ dependence.
These results correspond to $Q^2 \simeq0.5$ (GeV/$ c)^2$, $ W = 2.14 $ GeV, and $\theta_\mathrm{\gamma K}^\mathrm{c.m.} \simeq8 $ deg. We compare our results to past experiments and theoretical predictions in Figs. \ref{fig: CS_Q2depL2} and \ref{fig: CS_Q2depS2}. The results with the full dataset are shown in red and those with the divided dataset are shown in blue. Statistical errors are represented by solid error bars, while systematic errors are depicted as dashed boxes.

\begin{table*}[tbp]
\centering
\caption{Summary of the obtained differential cross sections}
\label{tab: dcs_results}
\begin{tabular}{lll} \hline\hline
$\gamma^*p\to K^+\Lambda$ &  &  \\
Full & $0.426^{+0.024}_{-0.023}(\mathrm{Stat.})^{+0.022}_{-0.045}(\mathrm{Syst.})\ \mu\mathrm{b/sr}$ & at $Q^2=0.2-0.8\ (\mathrm{GeV}/c)^2$\\
Divided-1 & $0.554^{+0.033}_{-0.032}(\mathrm{Stat.})^{+0.035}_{-0.079}(\mathrm{Syst.})\ \mu\mathrm{b/sr}$ & at $Q^2=0.2-0.5\ (\mathrm{GeV}/c)^2$\\
Divided-2 & $0.338\pm0.022(\mathrm{Stat.})^{+0.022}_{-0.055}(\mathrm{Syst.})\ \mu\mathrm{b/sr}$ & at $Q^2=0.5-0.8\ (\mathrm{GeV}/c)^2$ \\\hline
$\gamma^*p\to K^+\Sigma^0$ &  &  \\
Full & $0.086^{+0.009}_{-0.008}(\mathrm{Stat.})^{+0.041}_{-0.022}(\mathrm{Syst.})\ \mu\mathrm{b/sr}$ & at $Q^2=0.2-0.8\ (\mathrm{GeV}/c)^2$\\
Divided-1 & $0.128\pm0.013(\mathrm{Stat.})^{+0.061}_{-0.033}(\mathrm{Syst.})\ \mu\mathrm{b/sr}$ & at $Q^2=0.2-0.5\ (\mathrm{GeV}/c)^2$\\
Divided-2 & $0.041\pm0.006(\mathrm{Stat.})^{+0.020}_{-0.010}(\mathrm{Syst.})\ \mu\mathrm{b/sr}$ & at $Q^2=0.5-0.8\ (\mathrm{GeV}/c)^2$\\\hline\hline
\end{tabular}
\end{table*}

For comparison, other experimental data and theoretical predictions based on the isobaric approach are shown in the figures.
It should be noted that the other data and the theoretical calculations correspond to unpolarized differential cross sections defined as
\begin{align}
\frac{\mathrm{d}\sigma_\mathrm{UL}}{\mathrm{d}\Omega_\mathrm{K}^\mathrm{c.m.}}&:=\frac{\mathrm{d}\sigma_\mathrm{T}}{\mathrm{d}\Omega_\mathrm{K}^\mathrm{c.m.}}+\varepsilon\frac{\mathrm{d}\sigma_\mathrm{L}}{\mathrm{d}\Omega_\mathrm{K}^\mathrm{c.m.}}
\end{align}
which is obtained by averaging Eq. (\ref{eq: dcs_decomp}) over $360$ deg with respect to $\phi_{\gamma \mathrm{K}}$.

The other experimental data correspond to different kinematics \cite{bebek_electroproduction_1977,brown_coincidence_1972,bebek_electroproduction_1974}: a scaling method was used following the procedure described by Mohring {\it et al.} \cite{mohring_separation_2003} to compare the various data points. 
All of the experimental data shown in the figures were taken at forward angles, $\theta_{\gamma \mathrm{K}}^\mathrm{c.m.}<15$ deg. Also, major isobaric models, BS1 \cite{skoupil_photoproduction_2016}, BS2 \cite{skoupil_photoproduction_2016}, BS3 \cite{skoupil_photo-_2018}, KM \cite{mart_evidence_1999,lee_quasifree_2001}, Saclay-Lyon (SL) \cite{david_electromagnetic_1996}, SLA \cite{mizutani_off-shell_1998}, H2 \cite{bydzovsky_kaon_2005}, and Williams-Ji-Cotanch (WJC) \cite{williams_hyperon_1992}, are shown for $W=2.14$ GeV, $\theta_{\gamma\mathrm{K}}^\mathrm{c.m.}=8$ deg, and $\varepsilon=0.8$. The KM model is displayed up to its maximum computational range, $Q^2=2.2\ (\mathrm{GeV}/c)^2$.
Our results do not deviate significantly from the existing experimental data and the  theoretical calculations.

\section{Discussion\label{sec: Discussion}}

\begin{figure*}
\begin{minipage}[t]{0.48\textwidth}
\centering
\includegraphics[width=90mm]{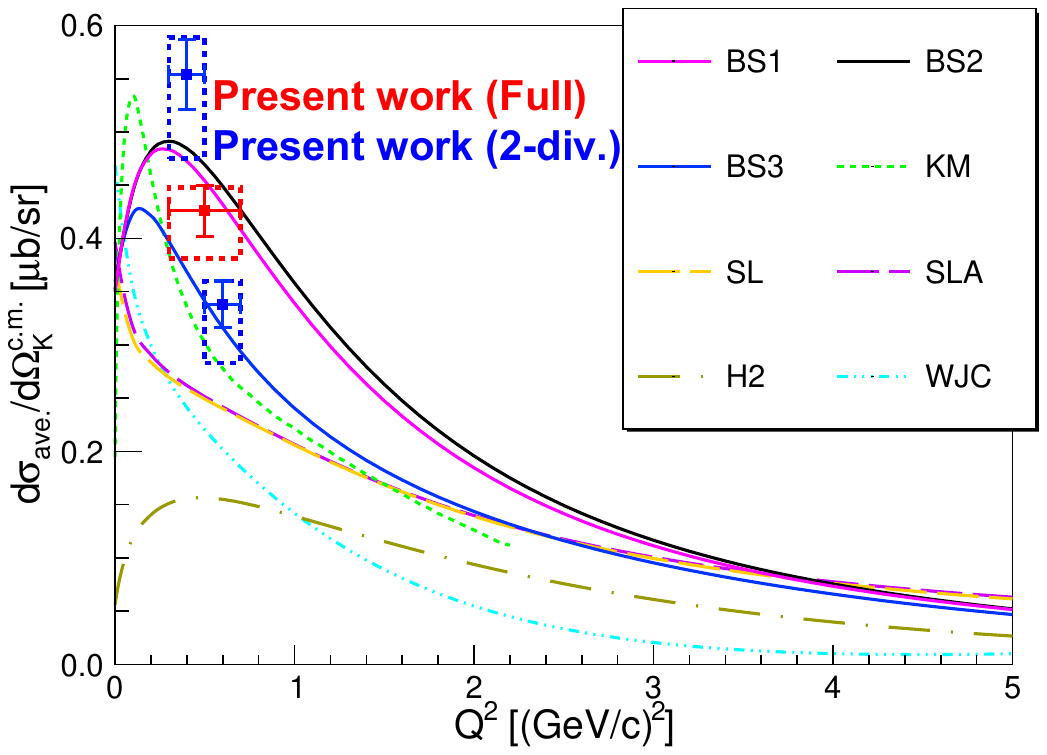}
\caption{
$Q^2$ dependence of the average differential cross section for the $p(\gamma^*,K^+)\Lambda$ reaction. Comparison in $Q^2$ dependence between our results and the isobaric models is shown.}
\label{fig: L_iso_Q2_tot_data}
\end{minipage}
\hfill
\begin{minipage}[t]{0.48\textwidth}
\centering
\includegraphics[width=90mm]{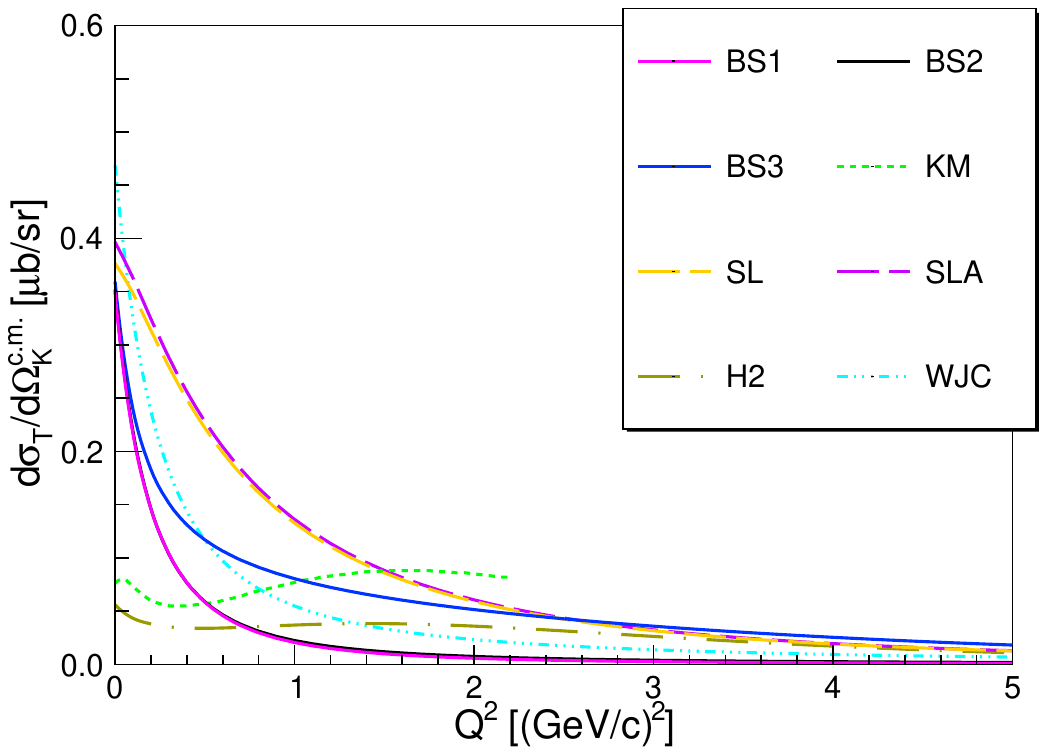}
\centering
\caption[$Q^2$ dependence of the transverse differential cross section for the $p(\gamma^*,K^+)\Lambda$ reaction]{
$Q^2$ dependence of the transverse differential cross section for the $p(\gamma^*,K^+)\Lambda$ reaction.}
\label{fig: L_iso_Q2_T}
\end{minipage}
\begin{minipage}[t]{0.48\textwidth}
\centering
\includegraphics[width=90mm]{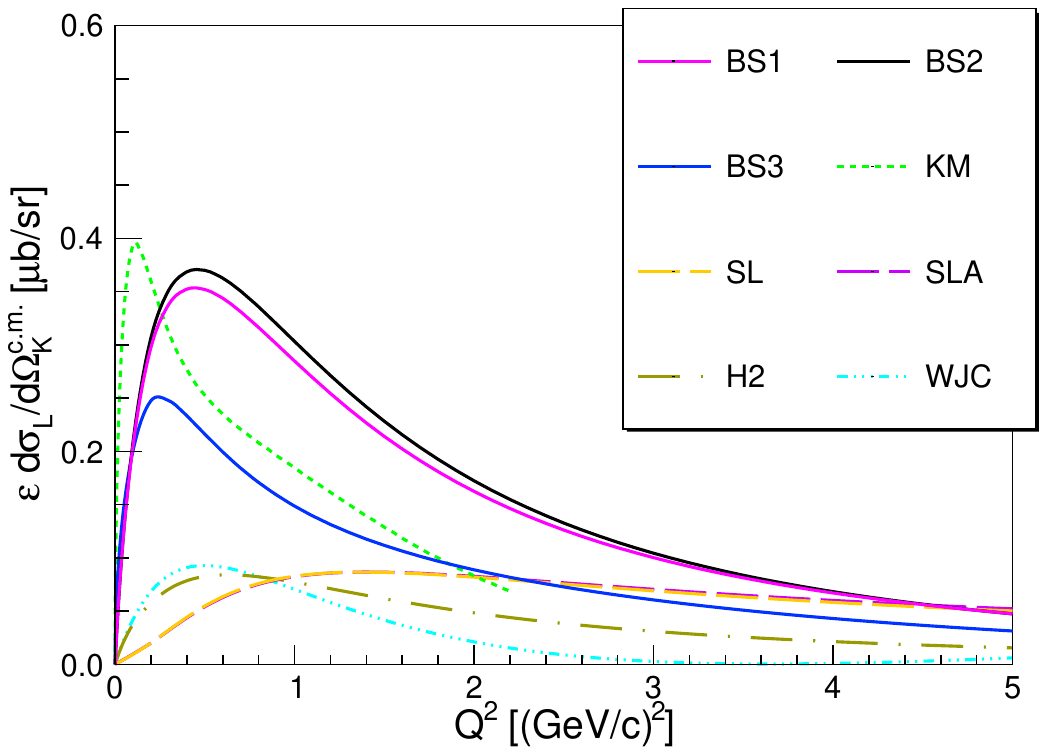}
\caption[$Q^2$ dependence of the longitudinal differential cross section for the $p(\gamma^*,K^+)\Lambda$ reaction]{
$Q^2$ dependence of the longitudinal differential cross section for the $p(\gamma^*,K^+)\Lambda$ reaction.}
\label{fig: L_iso_Q2_L}
\end{minipage}
\hfill
\begin{minipage}[t]{0.48\textwidth}
\centering
\includegraphics[width=90mm]{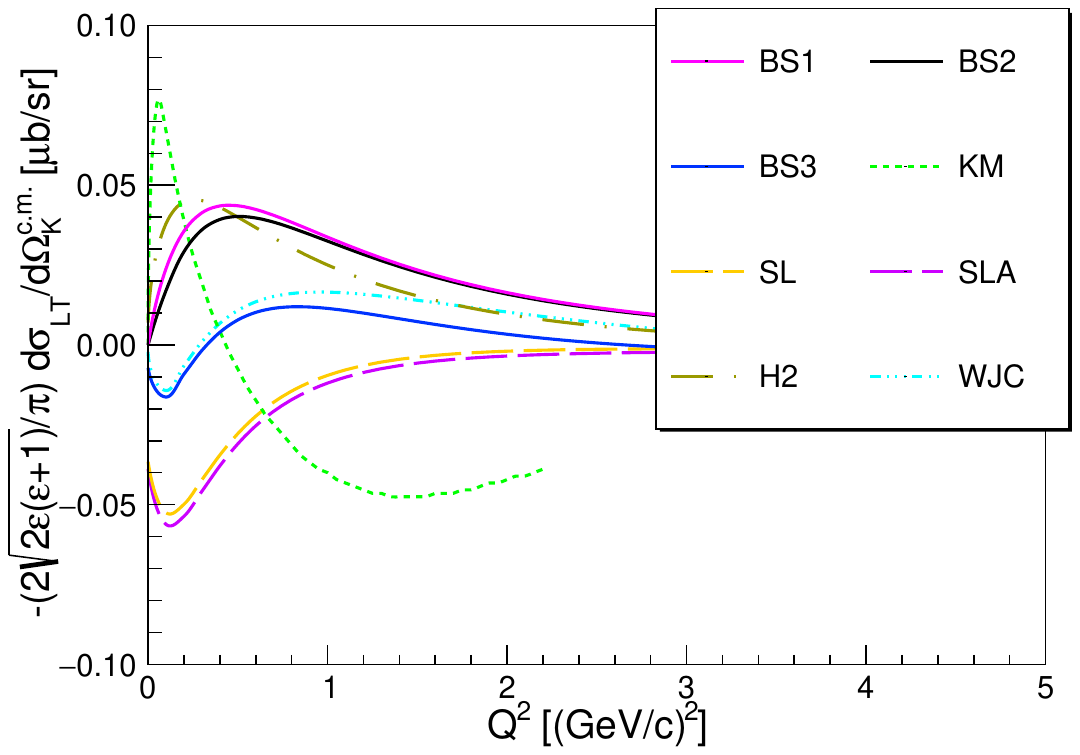}
\centering
\caption[$Q^2$ dependence of the transverse-longitudinal differential cross section for the $p(\gamma^*,K^+)\Lambda$ reaction]{
$Q^2$ dependence of the transverse-longitudinal differential cross section for the $p(\gamma^*,K^+)\Lambda$ reaction.}
\label{fig: L_iso_Q2_LT}
\end{minipage}
\end{figure*}
The unpolarized differential cross section is valid only if the acceptance covers the whole azimuthal angular range of $360$ deg. However, our experimental apparatus covered an angle approximately from $90$ deg to $270$ deg. Therefore, in our dataset, averaging over $\phi_{\gamma \mathrm{K}}$ within our experimental acceptance cancels the contribution of $\mathrm{d}\sigma_\mathrm{TT}$ but retains that of $\mathrm{d}\sigma_\mathrm{LT}$ with a numerical factor:
\begin{align}
\frac{\mathrm{d}\sigma_\mathrm{ave.}}{\mathrm{d}\Omega_\mathrm{K}^\mathrm{c.m.}}&:=\frac{\mathrm{d}\sigma_\mathrm{T}}{\mathrm{d}\Omega_\mathrm{K}^\mathrm{c.m.}}+\varepsilon\frac{\mathrm{d}\sigma_\mathrm{L}}{\mathrm{d}\Omega_\mathrm{K}^\mathrm{c.m.}}\nonumber\\
&\quad-\frac{2}{\pi}\sqrt{2\varepsilon\left(\varepsilon+1\right)}\frac{\mathrm{d}\sigma_\mathrm{LT}}{\mathrm{d}\Omega_\mathrm{K}^\mathrm{c.m.}}.\label{eq: dcs_decomp_cont}
\end{align}
This average differential cross section corresponds to the obtained results in Table \ref{tab: dcs_results}.
A comparison with predictions of the isobaric models calculated according to Eq. (\ref{eq: dcs_decomp_cont}) is shown in Fig. \ref{fig: L_iso_Q2_tot_data} for $\Lambda$ electroproduction. Similarly, contributions from the separate terms, the transverse $\mathrm{d} \sigma_\mathrm{T}$, longitudinal $\mathrm{d} \sigma_\mathrm{L}$, and longitudinal-transverse interference $\mathrm{d} \sigma_\mathrm{LT}$, are compared for the isobaric models in Figs. \ref{fig: L_iso_Q2_T}, \ref{fig: L_iso_Q2_L}, and \ref{fig: L_iso_Q2_LT}, respectively.

The average differential cross sections obtained in this experiment are reproduced by the BS1 and BS2 models \cite{skoupil_photoproduction_2016} as seen in Fig. \ref{fig: L_iso_Q2_tot_data}. These models were developed recently using also the new data from the CLAS \cite{mccracken_differential_2010} and LEPS \cite{sumihama__2006} collaborations and they were proven to reproduce well the photoproduction data at forward angles \cite{skoupil_photoproduction_2016}. Based on our new results, these models also reproduce the electroproduction data in our kinematic region. The BS3 model also reproduces our results relatively well. It is an extended version of the BS1 and BS2 models to electroproduction by adding couplings of the nucleon resonances to the longitudinal component of the virtual photon \cite{skoupil_photo-_2018}. This is also why the BS3 model predicts different $Q^2$ dependence of the longitudinal terms, $\mathrm{d} \sigma_\mathrm{L}$ and $\mathrm{d} \sigma_\mathrm{LT}$, observed in Figs. \ref{fig: L_iso_Q2_L} and \ref{fig: L_iso_Q2_LT}.

The KM model \cite{mart_evidence_1999,lee_quasifree_2001} has a steep rise at the average differential cross section at low $Q^2$ due to its strong $\mathrm{d}\sigma_\mathrm{L}$ term as shown in Fig. \ref{fig: L_iso_Q2_L}, and seems to reproduce our results.
However, the behavior in the low-$Q^2$ region was found to be inconsistent with the new electroproduction data of MAMI at $Q^2=0.055\ (\mathrm{GeV}/c)^2$ \cite{a1_collaboration_exclusive_2012}. In particular, the longitudinal component was shown to be too large below $Q^2=0.055\ (\mathrm{GeV}/c)^2$. While the KM is known to reproduce well photoproduction data, there is deviation from experimental data for electroproduction especially at low $Q^2$. This discrepancy is due to the poor knowledge of the longitudinal couplings since experimental data were scarce in the 1990s when it was developed. 
Furthermore, the number of resonances used in the KM model is only moderate similarly as in the older models like SL, SLA, H2, and WJC. In the $K\Lambda$ channel, the KM model utilizes only the $S_{11}(1650)$, $P_{11}(1710)$, $P_{13}(1720)$, and missing $D_{13}(1895)$ nucleon resonances, whereas in the $K\Sigma$ channels, the KM model makes use of only the $S_{11}(1650)$, $P_{11}(1710)$, $P_{13}(1720)$, $S_{31}(1900)$, and $P_{31}(1910)$ states. Therefore, in the $K\Lambda$ channel, contribution to the longitudinal terms from other resonances seems to be important to describe the small discrepancy between the KM model and the present data as indicated in Figs. \ref{fig: L_iso_Q2_L} and \ref{fig: L_iso_Q2_LT}. 

Next, the SL model \cite{david_electromagnetic_1996} and the SLA model \cite{mizutani_off-shell_1998} also can reproduce our results to some extent.
Note that predictions of these Saclay-Lyon models shown in the figures almost overlap each other. 
Even though these models do not include any couplings of the nucleon resonances with the longitudinal component of the virtual photon and they could not use the recent experimental data in their construction, they provide results which are close to ours in this kinematic condition.

Finally, the H2 \cite{bydzovsky_kaon_2005} and WJC model \cite{williams_hyperon_1992} are also relatively old models but their results were shown here for comparison. These models reveal similar $Q^2$ dependence as the other models but they do not reproduce the magnitude of the average differential cross section due to missing strength both in the transverse and longitudinal terms. The H2 model was fitted only to photoproduction data taken at CLAS in 2003 \cite{mcnabb_hyperon_2004}, and therefore it was not expected to show great results for electroproduction at other kinematic conditions.

The $Q^2$ dependence of the response functions in Eq. (\ref{eq: dcs_decomp_cont}) is determined by a competition between their genuine dependence on $Q^2$ stemming from dynamics of an isobaric model given by included Feynman diagrams \cite{skoupil_photoproduction_2016,skoupil_photo-_2018}, and dumping effects due to the electromagnetic form factors which mimic an internal structure of the hadrons included in the models.
In Figs. \ref{fig: L_iso_Q2_T} and \ref{fig: L_iso_Q2_L}, we show contributions from the transverse and the longitudinal terms in Eq. (\ref{eq: dcs_decomp_cont}), respectively. We see that the overall $Q^2$ dependence of the BS models observed in Fig. \ref{fig: L_iso_Q2_tot_data} at $Q^2<0.5\ (\mathrm{GeV}/c)^2$ is given by the longitudinal contribution. The peak observed in Fig. \ref{fig: L_iso_Q2_L} dominates the average differential cross sections and hence the BS models can reproduce our results at $Q^2\simeq0.5\ (\mathrm{GeV}/c)^2$. 
In contrast, the SL model predicts large values of $\mathrm{d}\sigma_\mathrm{T}$ but the $\mathrm{d} \sigma_\mathrm{L}$ gives only small contributions which makes the difference between the SL and BS models.

The longitudinal-transverse interference term is shown in Fig. \ref{fig: L_iso_Q2_LT}. This term has a relatively small effect on the average differential cross section, hence the similarities for the average differential cross section in Fig. \ref{fig: L_iso_Q2_tot_data} and unpolarized differential cross section in Fig. \ref{fig: CS_Q2depL2}.

Let us mention that the strong $Q^2$ dependence observed in the divided data
on the averaged full differential cross section can impose constraints on
dynamics of the theoretical models. Given the model dependence of this
discussion, the following discussion is based on the BS3 model as an example.
As already shown in Figs. \ref{fig: L_iso_Q2_T}, \ref{fig: L_iso_Q2_L}, and \ref{fig: L_iso_Q2_LT}, the separated contributions $\mathrm{d}\sigma_\mathrm{T}$, $\mathrm{d}\sigma_\mathrm{L}$, and $\mathrm{d}\sigma_\mathrm{LT}$ exhibit different $Q^2$ dependencies, with the peak structure being primarily formed by $\mathrm{d}\sigma_\mathrm{L}$. In the low-$Q^2$ region ($Q^2 < 0.5\ (\mathrm{GeV}/c)^2$), which includes our experimental condition, this peak is mainly formed due to the longitudinal couplings of the nucleon resonances rather than the electromagnetic form factors. Therefore, the pronounced $Q^2$ dependence observed in our experimental results is crucial for determining the longitudinal couplings
in the low-$Q^2$ region.
Significant sensitivity to a strength of the longitudinal couplings is also observed in $\mathrm{d}\sigma_\mathrm{LT}$ at $Q^2 < 1\ (\mathrm{GeV}/c)^2$  but there are almost no effects
in $\mathrm{d}\sigma_\mathrm{T}$ which reveals a steeply decreasing $Q^2$ dependence in the BS3 model at $Q^2 < 1\ (\mathrm{GeV}/c)^2$.
On the other hand, in the KM model, $\mathrm{d}\sigma_\mathrm{L}$ also shows a peak structure,
but $\mathrm{d}\sigma_\mathrm{T}$ remains flatter than that of the BS3, indicating different dynamics.
Nevertheless, as mentioned in Ref. \cite{a1_collaboration_exclusive_2012}, the longitudinal couplings play
a significant role in determining $\mathrm{d}\sigma_\mathrm{L}$ in the low-$Q^2$ region. Thus, our
experimental results, which reflect the $Q^2$ dependence in the low-$Q^2$ region,
can impose constraints on these theoretical models.
However, it is important to note that only two data points were obtained
in the experiment, each with a certain error.
There is also no need to emphasize that experimental results on the separated
$\mathrm{d}\sigma_\mathrm{T}$ and $\mathrm{d}\sigma_\mathrm{L}$ cross sections would allow one to draw more specific conclusions.

\begin{figure}[tb]
\centering
\includegraphics[width=90mm]{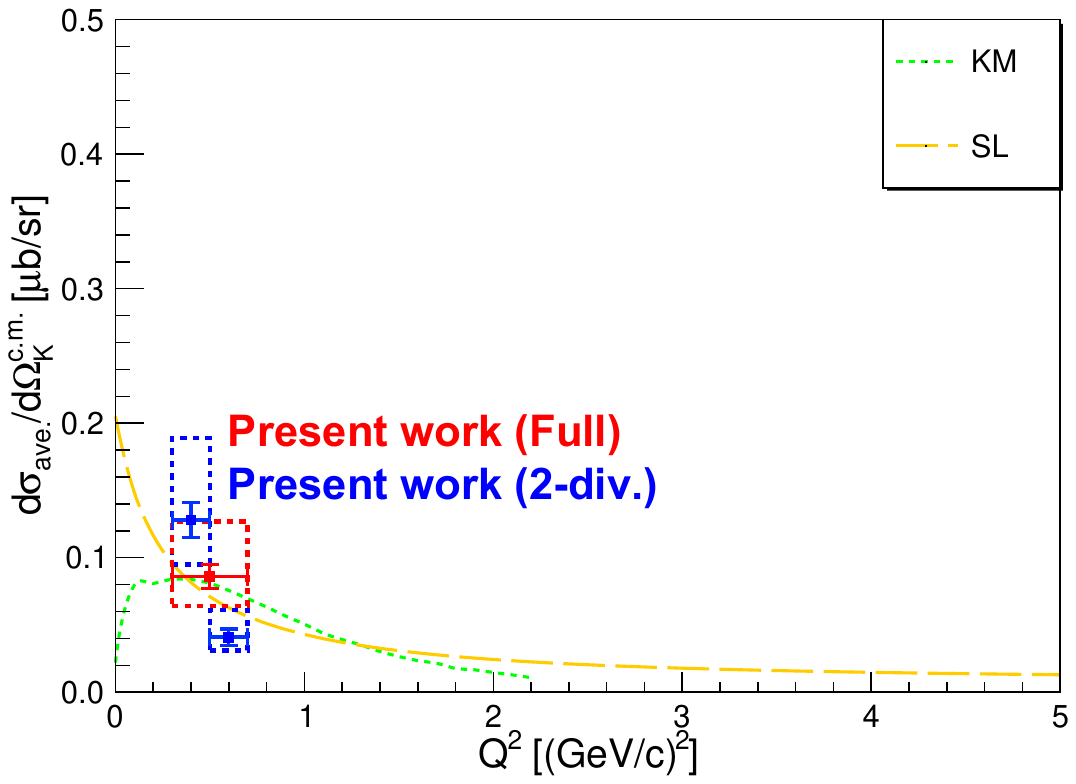}
\centering
\caption{
$Q^2$ dependence of the average differential cross section for the $p(\gamma^*,K^+)\Sigma^0$ reaction.}
\label{fig: S_iso_Q2_tot_data}
\end{figure}

For $\Sigma^0$ electroproduction, the KM model and the SL model are the only available isobaric models to compare with our results.
The average differential cross section for $\Sigma^0$ is shown in Fig. \ref{fig: S_iso_Q2_tot_data}. Similarly, the longitudinal-transverse interference term has a small contribution. Both the KM and SL models show similar results at our kinematic conditions. However, the SL model can reproduce our results of steep $Q^2$ dependence more accurately, although having large systematic errors due to the difficulties of the estimation of the radiative tail. For $\Sigma^0$ electroproduction, both experimental and theoretical updates are needed to clarify the differential cross section.

\section{Conclusion\label{sec: Summary}}
In the present paper, the differential cross sections for $\Lambda/\Sigma^0$ electroproduction at forward angles were reported. Despite the fact that there are abundant experimental data for photoproduction in a wide range of kinematics by CLAS, SAPHIR, LEPS, and GRAAL collaborations, understanding of production dynamics at forward angles has been limited because of the lack of experimental data at forward angles due to experimental difficulties. On the other hand, electroproduction data can be taken at forward angles but available data are still limited.

The results of the differential cross sections for $\Lambda/\Sigma^0$ electroproduction obtained in this experiment were compared with the theoretical predictions using the isobaric models. Our results provided new data at forward angles not covered by photoproduction so far. The best agreements of $\Lambda$ electroproduction are with the BS models, which are also in good agreement with the new results of CLAS. 

In the BS3 model, the longitudinal couplings of the virtual photon to nucleon fields play an important role in obtaining satisfactory results as shown in Ref. \cite{skoupil_photo-_2018}. At $Q^2\simeq0.5$ $(\mathrm{GeV}/c)^2$ contributions from these couplings enhance the longitudinal cross section, making a peak.
The contribution from the longitudinal couplings cannot be investigated by photoproduction, but can be approached only by electroproduction. Therefore our new data are an important input to the theoretical models in determining the magnitudes of the longitudinal couplings. This paper is expected not only to advance our comprehension of fundamental aspects of hyperon electroproduction and photoproduction but also to extend its relevance to applied studies, such as hypernuclear physics.

\appendix
%\section*{Acknowledgement}

\begin{acknowledgments}
We thank the JLab staff of the Division of Physics, Division of Accelerator, and the Division of Engineering for providing support for conducting the experiment.
This work was supported by U.S. Department of
Energy (DOE) grant DE-AC05-06OR23177 under which Jefferson Science Associates, LLC, operates the Thomas Jefferson National Accelerator Facility. The work of the Argonne National Laboratory group member is supported by DOE grant DE-AC02-06CH11357. The Kent State University contribution is supported under grant no. PHY-1714809 from the U.S. National Science Foundation. The hypernuclear program at JLab is supported by U.S. DOE grant DE-FG02-97ER41047. This work was supported by the Grant-in-Aid for Scientific Research on Innovative Areas “Toward new frontiers Encounter and synergy of state-of-the-art astronomical detectors and exotic quantum beams.” This work was supported by JSPS KAKENHI grants nos. 18H05459, 18H05457, 18H01219, 17H01121, 19J22055, 18H01220, 24H00219, and Grant-in-Aid for JSPS Research Fellow Grant No. 23KJ0100. This work was also supported by the Graduate Program on Physics for the Universe, Tohoku University (GP-PU), and SPIRITS 2020 of Kyoto University.
\end{acknowledgments}

\bibliography{elementary}

\begin{thebibliography}{32}%
\makeatletter
\providecommand \@ifxundefined [1]{%
 \@ifx{#1\undefined}
}%
\providecommand \@ifnum [1]{%
 \ifnum #1\expandafter \@firstoftwo
 \else \expandafter \@secondoftwo
 \fi
}%
\providecommand \@ifx [1]{%
 \ifx #1\expandafter \@firstoftwo
 \else \expandafter \@secondoftwo
 \fi
}%
\providecommand \natexlab [1]{#1}%
\providecommand \enquote  [1]{``#1''}%
\providecommand \bibnamefont  [1]{#1}%
\providecommand \bibfnamefont [1]{#1}%
\providecommand \citenamefont [1]{#1}%
\providecommand \href@noop [0]{\@secondoftwo}%
\providecommand \href [0]{\begingroup \@sanitize@url \@href}%
\providecommand \@href[1]{\@@startlink{#1}\@@href}%
\providecommand \@@href[1]{\endgroup#1\@@endlink}%
\providecommand \@sanitize@url [0]{\catcode `\\12\catcode `\$12\catcode
  `\&12\catcode `\#12\catcode `\^12\catcode `\_12\catcode `\%12\relax}%
\providecommand \@@startlink[1]{}%
\providecommand \@@endlink[0]{}%
\providecommand \url  [0]{\begingroup\@sanitize@url \@url }%
\providecommand \@url [1]{\endgroup\@href {#1}{\urlprefix }}%
\providecommand \urlprefix  [0]{URL }%
\providecommand \Eprint [0]{\href }%
\providecommand \doibase [0]{https://doi.org/}%
\providecommand \selectlanguage [0]{\@gobble}%
\providecommand \bibinfo  [0]{\@secondoftwo}%
\providecommand \bibfield  [0]{\@secondoftwo}%
\providecommand \translation [1]{[#1]}%
\providecommand \BibitemOpen [0]{}%
\providecommand \bibitemStop [0]{}%
\providecommand \bibitemNoStop [0]{.\EOS\space}%
\providecommand \EOS [0]{\spacefactor3000\relax}%
\providecommand \BibitemShut  [1]{\csname bibitem#1\endcsname}%
\let\auto@bib@innerbib\@empty
%</preamble>
\bibitem [{\citenamefont {Suzuki}\ \emph {et~al.}(2022)\citenamefont {Suzuki},
  \citenamefont {Gogami}, \citenamefont {Pandey}, \citenamefont {Itabashi},
  \citenamefont {Nagao}, \citenamefont {Okuyama}, \citenamefont {Nakamura},
  \citenamefont {Tang}, \citenamefont {Abrams}, \citenamefont {Akiyama},
  \citenamefont {Androic}, \citenamefont {Aniol}, \citenamefont {Gayoso},
  \citenamefont {Bane}, \citenamefont {Barcus} \emph
  {et~al.}}]{suzuki_cross-section_2022}%
  \BibitemOpen
  \bibfield  {author} {\bibinfo {author} {\bibfnamefont {K.~N.}\ \bibnamefont
  {Suzuki}}, \bibinfo {author} {\bibfnamefont {T.}~\bibnamefont {Gogami}},
  \bibinfo {author} {\bibfnamefont {B.}~\bibnamefont {Pandey}}, \bibinfo
  {author} {\bibfnamefont {K.}~\bibnamefont {Itabashi}}, \bibinfo {author}
  {\bibfnamefont {S.}~\bibnamefont {Nagao}}, \bibinfo {author} {\bibfnamefont
  {K.}~\bibnamefont {Okuyama}}, \bibinfo {author} {\bibfnamefont {S.~N.}\
  \bibnamefont {Nakamura}}, \bibinfo {author} {\bibfnamefont {L.}~\bibnamefont
  {Tang}}, \bibinfo {author} {\bibfnamefont {D.}~\bibnamefont {Abrams}},
  \bibinfo {author} {\bibfnamefont {T.}~\bibnamefont {Akiyama}}, \bibinfo
  {author} {\bibfnamefont {D.}~\bibnamefont {Androic}}, \bibinfo {author}
  {\bibfnamefont {K.}~\bibnamefont {Aniol}}, \bibinfo {author} {\bibfnamefont
  {C.~A.}\ \bibnamefont {Gayoso}}, \bibinfo {author} {\bibfnamefont
  {J.}~\bibnamefont {Bane}}, \bibinfo {author} {\bibfnamefont {S.}~\bibnamefont
  {Barcus}}, \emph {et~al.},\ }\href {https://doi.org/10.1093/ptep/ptab158}
  {\bibfield  {journal} {\bibinfo  {journal} {Prog. Theor. Exp. Phys.}\
  }\textbf {\bibinfo {volume} {2022}},\ \bibinfo {pages} {013D01} (\bibinfo
  {year} {2022})}\BibitemShut {NoStop}%
\bibitem [{\citenamefont {Pandey}\ \emph {et~al.}(2022)\citenamefont {Pandey},
  \citenamefont {Tang}, \citenamefont {Gogami}, \citenamefont {Suzuki},
  \citenamefont {Itabashi}, \citenamefont {Nagao}, \citenamefont {Okuyama},
  \citenamefont {Nakamura}, \citenamefont {Abrams}, \citenamefont {Afnan},
  \citenamefont {Akiyama}, \citenamefont {Androic}, \citenamefont {Aniol},
  \citenamefont {Averett}, \citenamefont {Ayerbe~Gayoso} \emph
  {et~al.}}]{pandey_spectroscopic_2022}%
  \BibitemOpen
  \bibfield  {author} {\bibinfo {author} {\bibfnamefont {B.}~\bibnamefont
  {Pandey}}, \bibinfo {author} {\bibfnamefont {L.}~\bibnamefont {Tang}},
  \bibinfo {author} {\bibfnamefont {T.}~\bibnamefont {Gogami}}, \bibinfo
  {author} {\bibfnamefont {K.~N.}\ \bibnamefont {Suzuki}}, \bibinfo {author}
  {\bibfnamefont {K.}~\bibnamefont {Itabashi}}, \bibinfo {author}
  {\bibfnamefont {S.}~\bibnamefont {Nagao}}, \bibinfo {author} {\bibfnamefont
  {K.}~\bibnamefont {Okuyama}}, \bibinfo {author} {\bibfnamefont {S.~N.}\
  \bibnamefont {Nakamura}}, \bibinfo {author} {\bibfnamefont {D.}~\bibnamefont
  {Abrams}}, \bibinfo {author} {\bibfnamefont {I.~R.}\ \bibnamefont {Afnan}},
  \bibinfo {author} {\bibfnamefont {T.}~\bibnamefont {Akiyama}}, \bibinfo
  {author} {\bibfnamefont {D.}~\bibnamefont {Androic}}, \bibinfo {author}
  {\bibfnamefont {K.}~\bibnamefont {Aniol}}, \bibinfo {author} {\bibfnamefont
  {T.}~\bibnamefont {Averett}}, \bibinfo {author} {\bibfnamefont
  {C.}~\bibnamefont {Ayerbe~Gayoso}}, \emph {et~al.},\ }\href
  {https://doi.org/10.1103/PhysRevC.105.L051001} {\bibfield  {journal}
  {\bibinfo  {journal} {Phys. Rev. C}\ }\textbf {\bibinfo {volume} {105}},\
  \bibinfo {pages} {L051001} (\bibinfo {year} {2022})}\BibitemShut {NoStop}%
\bibitem [{\citenamefont {Mart}\ and\ \citenamefont
  {Bennhold}(1999)}]{mart_evidence_1999}%
  \BibitemOpen
  \bibfield  {author} {\bibinfo {author} {\bibfnamefont {T.}~\bibnamefont
  {Mart}}\ and\ \bibinfo {author} {\bibfnamefont {C.}~\bibnamefont
  {Bennhold}},\ }\href {https://doi.org/10.1103/PhysRevC.61.012201} {\bibfield
  {journal} {\bibinfo  {journal} {Phys. Rev. C}\ }\textbf {\bibinfo {volume}
  {61}},\ \bibinfo {pages} {012201(R)} (\bibinfo {year} {1999})}\BibitemShut
  {NoStop}%
\bibitem [{\citenamefont {Lee}\ \emph {et~al.}(2001)\citenamefont {Lee},
  \citenamefont {Mart}, \citenamefont {Bennhold}, \citenamefont {Haberzettl},\
  and\ \citenamefont {Wright}}]{lee_quasifree_2001}%
  \BibitemOpen
  \bibfield  {author} {\bibinfo {author} {\bibfnamefont {F.}~\bibnamefont
  {Lee}}, \bibinfo {author} {\bibfnamefont {T.}~\bibnamefont {Mart}}, \bibinfo
  {author} {\bibfnamefont {C.}~\bibnamefont {Bennhold}}, \bibinfo {author}
  {\bibfnamefont {H.}~\bibnamefont {Haberzettl}},\ and\ \bibinfo {author}
  {\bibfnamefont {L.}~\bibnamefont {Wright}},\ }\href
  {https://doi.org/10.1016/S0375-9474(01)01098-3} {\bibfield  {journal}
  {\bibinfo  {journal} {Nucl. Phys. A}\ }\textbf {\bibinfo {volume} {695}},\
  \bibinfo {pages} {237} (\bibinfo {year} {2001})}\BibitemShut {NoStop}%
\bibitem [{\citenamefont {David}\ \emph {et~al.}(1996)\citenamefont {David},
  \citenamefont {Fayard}, \citenamefont {Lamot},\ and\ \citenamefont
  {Saghai}}]{david_electromagnetic_1996}%
  \BibitemOpen
  \bibfield  {author} {\bibinfo {author} {\bibfnamefont {J.~C.}\ \bibnamefont
  {David}}, \bibinfo {author} {\bibfnamefont {C.}~\bibnamefont {Fayard}},
  \bibinfo {author} {\bibfnamefont {G.~H.}\ \bibnamefont {Lamot}},\ and\
  \bibinfo {author} {\bibfnamefont {B.}~\bibnamefont {Saghai}},\ }\href
  {https://doi.org/10.1103/PhysRevC.53.2613} {\bibfield  {journal} {\bibinfo
  {journal} {Phys. Rev. C}\ }\textbf {\bibinfo {volume} {53}},\ \bibinfo
  {pages} {2613} (\bibinfo {year} {1996})}\BibitemShut {NoStop}%
\bibitem [{\citenamefont {Mizutani}\ \emph {et~al.}(1998)\citenamefont
  {Mizutani}, \citenamefont {Fayard}, \citenamefont {Lamot},\ and\
  \citenamefont {Saghai}}]{mizutani_off-shell_1998}%
  \BibitemOpen
  \bibfield  {author} {\bibinfo {author} {\bibfnamefont {T.}~\bibnamefont
  {Mizutani}}, \bibinfo {author} {\bibfnamefont {C.}~\bibnamefont {Fayard}},
  \bibinfo {author} {\bibfnamefont {G.-H.}\ \bibnamefont {Lamot}},\ and\
  \bibinfo {author} {\bibfnamefont {B.}~\bibnamefont {Saghai}},\ }\href
  {https://doi.org/10.1103/PhysRevC.58.75} {\bibfield  {journal} {\bibinfo
  {journal} {Phys. Rev. C}\ }\textbf {\bibinfo {volume} {58}},\ \bibinfo
  {pages} {75} (\bibinfo {year} {1998})}\BibitemShut {NoStop}%
\bibitem [{\citenamefont {Byd^^c5^^beovsk^^c3^^bd}\ \emph
  {et~al.}(2003)\citenamefont {Byd^^c5^^beovsk^^c3^^bd}, \citenamefont
  {Cusanno}, \citenamefont {Frullani}, \citenamefont {Garibaldi}, \citenamefont
  {Iodice}, \citenamefont {Sotona},\ and\ \citenamefont
  {Urciuoli}}]{bydzovsky_models_2003}%
  \BibitemOpen
  \bibfield  {author} {\bibinfo {author} {\bibfnamefont {P.}~\bibnamefont
  {Byd^^c5^^beovsk^^c3^^bd}}, \bibinfo {author} {\bibfnamefont
  {F.}~\bibnamefont {Cusanno}}, \bibinfo {author} {\bibfnamefont
  {S.}~\bibnamefont {Frullani}}, \bibinfo {author} {\bibfnamefont
  {F.}~\bibnamefont {Garibaldi}}, \bibinfo {author} {\bibfnamefont
  {M.}~\bibnamefont {Iodice}}, \bibinfo {author} {\bibfnamefont
  {M.}~\bibnamefont {Sotona}},\ and\ \bibinfo {author} {\bibfnamefont {G.~M.}\
  \bibnamefont {Urciuoli}},\ }\href {http://arxiv.org/abs/nucl-th/0305039}
  {\bibfield  {journal} {\bibinfo  {journal} {arXiv:nucl-th/0305039}\ }
  (\bibinfo {year} {2003})}\BibitemShut {NoStop}%
\bibitem [{\citenamefont {Skoupil}\ and\ \citenamefont
  {Byd^^c5^^beovsk^^c3^^bd}(2016)}]{skoupil_photoproduction_2016}%
  \BibitemOpen
  \bibfield  {author} {\bibinfo {author} {\bibfnamefont {D.}~\bibnamefont
  {Skoupil}}\ and\ \bibinfo {author} {\bibfnamefont {P.}~\bibnamefont
  {Byd^^c5^^beovsk^^c3^^bd}},\ }\href
  {https://doi.org/10.1103/PhysRevC.93.025204} {\bibfield  {journal} {\bibinfo
  {journal} {Phys. Rev. C}\ }\textbf {\bibinfo {volume} {93}},\ \bibinfo
  {pages} {025204} (\bibinfo {year} {2016})}\BibitemShut {NoStop}%
\bibitem [{\citenamefont {Skoupil}\ and\ \citenamefont
  {Byd^^c5^^beovsk^^c3^^bd}(2018)}]{skoupil_photo-_2018}%
  \BibitemOpen
  \bibfield  {author} {\bibinfo {author} {\bibfnamefont {D.}~\bibnamefont
  {Skoupil}}\ and\ \bibinfo {author} {\bibfnamefont {P.}~\bibnamefont
  {Byd^^c5^^beovsk^^c3^^bd}},\ }\href
  {https://doi.org/10.1103/PhysRevC.97.025202} {\bibfield  {journal} {\bibinfo
  {journal} {Phys. Rev. C}\ }\textbf {\bibinfo {volume} {97}},\ \bibinfo
  {pages} {025202} (\bibinfo {year} {2018})}\BibitemShut {NoStop}%
\bibitem [{\citenamefont {Byd^^c5^^beovsk^^c3^^bd}\ and\ \citenamefont
  {Sotona}(2005)}]{bydzovsky_kaon_2005}%
  \BibitemOpen
  \bibfield  {author} {\bibinfo {author} {\bibfnamefont {P.}~\bibnamefont
  {Byd^^c5^^beovsk^^c3^^bd}}\ and\ \bibinfo {author} {\bibfnamefont
  {M.}~\bibnamefont {Sotona}},\ }\href
  {https://doi.org/10.1016/j.nuclphysa.2005.02.076} {\bibfield  {journal}
  {\bibinfo  {journal} {Nucl. Phys. A}\ }\textbf {\bibinfo {volume} {754}},\
  \bibinfo {pages} {243c} (\bibinfo {year} {2005})}\BibitemShut {NoStop}%
\bibitem [{\citenamefont {Williams}\ \emph {et~al.}(1992)\citenamefont
  {Williams}, \citenamefont {Ji},\ and\ \citenamefont
  {Cotanch}}]{williams_hyperon_1992}%
  \BibitemOpen
  \bibfield  {author} {\bibinfo {author} {\bibfnamefont {R.~A.}\ \bibnamefont
  {Williams}}, \bibinfo {author} {\bibfnamefont {C.-R.}\ \bibnamefont {Ji}},\
  and\ \bibinfo {author} {\bibfnamefont {S.~R.}\ \bibnamefont {Cotanch}},\
  }\href {https://doi.org/10.1103/PhysRevC.46.1617} {\bibfield  {journal}
  {\bibinfo  {journal} {Phys. Rev. C}\ }\textbf {\bibinfo {volume} {46}},\
  \bibinfo {pages} {1617} (\bibinfo {year} {1992})}\BibitemShut {NoStop}%
\bibitem [{\citenamefont {Corthals}\ \emph {et~al.}(2006)\citenamefont
  {Corthals}, \citenamefont {Ryckebusch},\ and\ \citenamefont
  {Cauteren}}]{corthals_forward-angle_2006}%
  \BibitemOpen
  \bibfield  {author} {\bibinfo {author} {\bibfnamefont {T.}~\bibnamefont
  {Corthals}}, \bibinfo {author} {\bibfnamefont {J.}~\bibnamefont
  {Ryckebusch}},\ and\ \bibinfo {author} {\bibfnamefont {T.~V.}\ \bibnamefont
  {Cauteren}},\ }\href {https://doi.org/10.1103/PhysRevC.73.045207} {\bibfield
  {journal} {\bibinfo  {journal} {Phys. Rev. C}\ }\textbf {\bibinfo {volume}
  {73}},\ \bibinfo {pages} {045207} (\bibinfo {year} {2006})}\BibitemShut
  {NoStop}%
\bibitem [{\citenamefont {De^^c2^^a0Cruz}\ \emph {et~al.}(2012)\citenamefont
  {De^^c2^^a0Cruz}, \citenamefont {Vrancx}, \citenamefont {Vancraeyveld},\ and\
  \citenamefont {Ryckebusch}}]{decruz_bayesian_2012}%
  \BibitemOpen
  \bibfield  {author} {\bibinfo {author} {\bibfnamefont {L.}~\bibnamefont
  {De^^c2^^a0Cruz}}, \bibinfo {author} {\bibfnamefont {T.}~\bibnamefont
  {Vrancx}}, \bibinfo {author} {\bibfnamefont {P.}~\bibnamefont
  {Vancraeyveld}},\ and\ \bibinfo {author} {\bibfnamefont {J.}~\bibnamefont
  {Ryckebusch}},\ }\href {https://doi.org/10.1103/PhysRevLett.108.182002}
  {\bibfield  {journal} {\bibinfo  {journal} {Phys. Rev. Lett.}\ }\textbf
  {\bibinfo {volume} {108}},\ \bibinfo {pages} {182002} (\bibinfo {year}
  {2012})}\BibitemShut {NoStop}%
\bibitem [{\citenamefont {Byd^^c5^^beovsk^^c3^^bd}\ and\ \citenamefont
  {Skoupil}(2019)}]{bydzovsky_photoproduction_2019}%
  \BibitemOpen
  \bibfield  {author} {\bibinfo {author} {\bibfnamefont {P.}~\bibnamefont
  {Byd^^c5^^beovsk^^c3^^bd}}\ and\ \bibinfo {author} {\bibfnamefont
  {D.}~\bibnamefont {Skoupil}},\ }\href
  {https://doi.org/10.1103/PhysRevC.100.035202} {\bibfield  {journal} {\bibinfo
   {journal} {Phys. Rev. C}\ }\textbf {\bibinfo {volume} {100}},\ \bibinfo
  {pages} {035202} (\bibinfo {year} {2019})}\BibitemShut {NoStop}%
\bibitem [{\citenamefont {Amaldi}\ \emph {et~al.}(1979)\citenamefont {Amaldi},
  \citenamefont {Fubini},\ and\ \citenamefont {Furlan}}]{amaldi_pion_1979}%
  \BibitemOpen
  \bibfield  {author} {\bibinfo {author} {\bibfnamefont {E.}~\bibnamefont
  {Amaldi}}, \bibinfo {author} {\bibfnamefont {S.}~\bibnamefont {Fubini}},\
  and\ \bibinfo {author} {\bibfnamefont {G.}~\bibnamefont {Furlan}},\
  }\href@noop {} {\emph {\bibinfo {title} {Pion {Electroproduction}.
  {Electroproduction} at {Low}- {Energy} and {Hadron} {Form}-{Factors}}}},\
  Vol.~\bibinfo {volume} {83}\ (\bibinfo  {publisher} {Springer Tracts in
  Modern Physics},\ \bibinfo {year} {1979})\BibitemShut {NoStop}%
\bibitem [{\citenamefont {Bradford}\ \emph {et~al.}(2006)\citenamefont
  {Bradford}, \citenamefont {Schumacher}, \citenamefont {McNabb}, \citenamefont
  {Todor}, \citenamefont {Adams}, \citenamefont {Ambrozewicz}, \citenamefont
  {Anciant}, \citenamefont {Anghinolfi}, \citenamefont {Asavapibhop},
  \citenamefont {Asryan}, \citenamefont {Audit}, \citenamefont {Avakian},
  \citenamefont {Bagdasaryan}, \citenamefont {Baillie}, \citenamefont {Ball}
  \emph {et~al.}}]{bradford_differential_2006}%
  \BibitemOpen
  \bibfield  {author} {\bibinfo {author} {\bibfnamefont {R.}~\bibnamefont
  {Bradford}}, \bibinfo {author} {\bibfnamefont {R.~A.}\ \bibnamefont
  {Schumacher}}, \bibinfo {author} {\bibfnamefont {J.~W.~C.}\ \bibnamefont
  {McNabb}}, \bibinfo {author} {\bibfnamefont {L.}~\bibnamefont {Todor}},
  \bibinfo {author} {\bibfnamefont {G.}~\bibnamefont {Adams}}, \bibinfo
  {author} {\bibfnamefont {P.}~\bibnamefont {Ambrozewicz}}, \bibinfo {author}
  {\bibfnamefont {E.}~\bibnamefont {Anciant}}, \bibinfo {author} {\bibfnamefont
  {M.}~\bibnamefont {Anghinolfi}}, \bibinfo {author} {\bibfnamefont
  {B.}~\bibnamefont {Asavapibhop}}, \bibinfo {author} {\bibfnamefont
  {G.}~\bibnamefont {Asryan}}, \bibinfo {author} {\bibfnamefont
  {G.}~\bibnamefont {Audit}}, \bibinfo {author} {\bibfnamefont
  {H.}~\bibnamefont {Avakian}}, \bibinfo {author} {\bibfnamefont
  {H.}~\bibnamefont {Bagdasaryan}}, \bibinfo {author} {\bibfnamefont
  {N.}~\bibnamefont {Baillie}}, \bibinfo {author} {\bibfnamefont {J.~P.}\
  \bibnamefont {Ball}}, \emph {et~al.},\ }\href
  {https://doi.org/10.1103/PhysRevC.73.035202} {\bibfield  {journal} {\bibinfo
  {journal} {Phys. Rev. C}\ }\textbf {\bibinfo {volume} {73}},\ \bibinfo
  {pages} {035202} (\bibinfo {year} {2006})}\BibitemShut {NoStop}%
\bibitem [{\citenamefont {McCracken}\ \emph {et~al.}(2010)\citenamefont
  {McCracken}, \citenamefont {Bellis}, \citenamefont {Meyer}, \citenamefont
  {Williams}, \citenamefont {Adhikari}, \citenamefont {Anghinolfi},
  \citenamefont {Ball}, \citenamefont {Battaglieri}, \citenamefont {Berman},
  \citenamefont {Biselli}, \citenamefont {Branford}, \citenamefont {Briscoe},
  \citenamefont {Brooks}, \citenamefont {Burkert}, \citenamefont {Careccia}
  \emph {et~al.}}]{mccracken_differential_2010}%
  \BibitemOpen
  \bibfield  {author} {\bibinfo {author} {\bibfnamefont {M.~E.}\ \bibnamefont
  {McCracken}}, \bibinfo {author} {\bibfnamefont {M.}~\bibnamefont {Bellis}},
  \bibinfo {author} {\bibfnamefont {C.~A.}\ \bibnamefont {Meyer}}, \bibinfo
  {author} {\bibfnamefont {M.}~\bibnamefont {Williams}}, \bibinfo {author}
  {\bibfnamefont {K.~P.}\ \bibnamefont {Adhikari}}, \bibinfo {author}
  {\bibfnamefont {M.}~\bibnamefont {Anghinolfi}}, \bibinfo {author}
  {\bibfnamefont {J.}~\bibnamefont {Ball}}, \bibinfo {author} {\bibfnamefont
  {M.}~\bibnamefont {Battaglieri}}, \bibinfo {author} {\bibfnamefont {B.~L.}\
  \bibnamefont {Berman}}, \bibinfo {author} {\bibfnamefont {A.~S.}\
  \bibnamefont {Biselli}}, \bibinfo {author} {\bibfnamefont {D.}~\bibnamefont
  {Branford}}, \bibinfo {author} {\bibfnamefont {W.~J.}\ \bibnamefont
  {Briscoe}}, \bibinfo {author} {\bibfnamefont {W.~K.}\ \bibnamefont {Brooks}},
  \bibinfo {author} {\bibfnamefont {V.~D.}\ \bibnamefont {Burkert}}, \bibinfo
  {author} {\bibfnamefont {S.~L.}\ \bibnamefont {Careccia}}, \emph {et~al.},\
  }\href {https://doi.org/10.1103/PhysRevC.81.025201} {\bibfield  {journal}
  {\bibinfo  {journal} {Phys. Rev. C}\ }\textbf {\bibinfo {volume} {81}},\
  \bibinfo {pages} {025201} (\bibinfo {year} {2010})}\BibitemShut {NoStop}%
\bibitem [{\citenamefont {McNabb}\ \emph {et~al.}(2004)\citenamefont {McNabb},
  \citenamefont {Schumacher}, \citenamefont {Todor}, \citenamefont {Adams},
  \citenamefont {Anciant}, \citenamefont {Anghinolfi}, \citenamefont
  {Asavapibhop}, \citenamefont {Audit}, \citenamefont {Auger}, \citenamefont
  {Avakian}, \citenamefont {Bagdasaryan}, \citenamefont {Ball}, \citenamefont
  {Barrow}, \citenamefont {Battaglieri}, \citenamefont {Beard} \emph
  {et~al.}}]{mcnabb_hyperon_2004}%
  \BibitemOpen
  \bibfield  {author} {\bibinfo {author} {\bibfnamefont {J.~W.~C.}\
  \bibnamefont {McNabb}}, \bibinfo {author} {\bibfnamefont {R.~A.}\
  \bibnamefont {Schumacher}}, \bibinfo {author} {\bibfnamefont
  {L.}~\bibnamefont {Todor}}, \bibinfo {author} {\bibfnamefont
  {G.}~\bibnamefont {Adams}}, \bibinfo {author} {\bibfnamefont
  {E.}~\bibnamefont {Anciant}}, \bibinfo {author} {\bibfnamefont
  {M.}~\bibnamefont {Anghinolfi}}, \bibinfo {author} {\bibfnamefont
  {B.}~\bibnamefont {Asavapibhop}}, \bibinfo {author} {\bibfnamefont
  {G.}~\bibnamefont {Audit}}, \bibinfo {author} {\bibfnamefont
  {T.}~\bibnamefont {Auger}}, \bibinfo {author} {\bibfnamefont
  {H.}~\bibnamefont {Avakian}}, \bibinfo {author} {\bibfnamefont
  {H.}~\bibnamefont {Bagdasaryan}}, \bibinfo {author} {\bibfnamefont {J.~P.}\
  \bibnamefont {Ball}}, \bibinfo {author} {\bibfnamefont {S.}~\bibnamefont
  {Barrow}}, \bibinfo {author} {\bibfnamefont {M.}~\bibnamefont {Battaglieri}},
  \bibinfo {author} {\bibfnamefont {K.}~\bibnamefont {Beard}}, \emph {et~al.},\
  }\href {https://doi.org/10.1103/PhysRevC.69.042201} {\bibfield  {journal}
  {\bibinfo  {journal} {Phys. Rev. C}\ }\textbf {\bibinfo {volume} {69}},\
  \bibinfo {pages} {042201} (\bibinfo {year} {2004})}\BibitemShut {NoStop}%
\bibitem [{\citenamefont {Dey}\ \emph {et~al.}(2010)\citenamefont {Dey},
  \citenamefont {Meyer}, \citenamefont {Bellis}, \citenamefont {McCracken},
  \citenamefont {Williams}, \citenamefont {Adhikari}, \citenamefont {Aghasyan},
  \citenamefont {Anghinolfi}, \citenamefont {Ball}, \citenamefont
  {Battaglieri}, \citenamefont {Batourine}, \citenamefont {Bedlinskiy},
  \citenamefont {Berman}, \citenamefont {Biselli}, \citenamefont {Branford}
  \emph {et~al.}}]{dey_differential_2010}%
  \BibitemOpen
  \bibfield  {author} {\bibinfo {author} {\bibfnamefont {B.}~\bibnamefont
  {Dey}}, \bibinfo {author} {\bibfnamefont {C.~A.}\ \bibnamefont {Meyer}},
  \bibinfo {author} {\bibfnamefont {M.}~\bibnamefont {Bellis}}, \bibinfo
  {author} {\bibfnamefont {M.~E.}\ \bibnamefont {McCracken}}, \bibinfo {author}
  {\bibfnamefont {M.}~\bibnamefont {Williams}}, \bibinfo {author}
  {\bibfnamefont {K.~P.}\ \bibnamefont {Adhikari}}, \bibinfo {author}
  {\bibfnamefont {M.}~\bibnamefont {Aghasyan}}, \bibinfo {author}
  {\bibfnamefont {M.}~\bibnamefont {Anghinolfi}}, \bibinfo {author}
  {\bibfnamefont {J.}~\bibnamefont {Ball}}, \bibinfo {author} {\bibfnamefont
  {M.}~\bibnamefont {Battaglieri}}, \bibinfo {author} {\bibfnamefont
  {V.}~\bibnamefont {Batourine}}, \bibinfo {author} {\bibfnamefont
  {I.}~\bibnamefont {Bedlinskiy}}, \bibinfo {author} {\bibfnamefont {B.~L.}\
  \bibnamefont {Berman}}, \bibinfo {author} {\bibfnamefont {A.~S.}\
  \bibnamefont {Biselli}}, \bibinfo {author} {\bibfnamefont {D.}~\bibnamefont
  {Branford}}, \emph {et~al.},\ }\href
  {https://doi.org/10.1103/PhysRevC.82.025202} {\bibfield  {journal} {\bibinfo
  {journal} {Phys. Rev. C}\ }\textbf {\bibinfo {volume} {82}},\ \bibinfo
  {pages} {025202} (\bibinfo {year} {2010})}\BibitemShut {NoStop}%
\bibitem [{\citenamefont {Glander}\ \emph {et~al.}(2004)\citenamefont
  {Glander}, \citenamefont {Barth}, \citenamefont {Braun}, \citenamefont
  {Hannappel}, \citenamefont {J^^c3^^b6pen}, \citenamefont {Klein},
  \citenamefont {Klempt}, \citenamefont {Lawall}, \citenamefont {Link},
  \citenamefont {Menze}, \citenamefont {Neuerburg}, \citenamefont {Ostrick},
  \citenamefont {Paul}, \citenamefont {Schulday}, \citenamefont {Schwille},
  \citenamefont {Pee}, \citenamefont {Wieland}, \citenamefont
  {Wi^^c3^^9fkirchen},\ and\ \citenamefont {Wu}}]{glander_measurement_2004}%
  \BibitemOpen
  \bibfield  {author} {\bibinfo {author} {\bibfnamefont {K.-H.}\ \bibnamefont
  {Glander}}, \bibinfo {author} {\bibfnamefont {J.}~\bibnamefont {Barth}},
  \bibinfo {author} {\bibfnamefont {W.}~\bibnamefont {Braun}}, \bibinfo
  {author} {\bibfnamefont {J.}~\bibnamefont {Hannappel}}, \bibinfo {author}
  {\bibfnamefont {N.}~\bibnamefont {J^^c3^^b6pen}}, \bibinfo {author}
  {\bibfnamefont {F.}~\bibnamefont {Klein}}, \bibinfo {author} {\bibfnamefont
  {E.}~\bibnamefont {Klempt}}, \bibinfo {author} {\bibfnamefont
  {R.}~\bibnamefont {Lawall}}, \bibinfo {author} {\bibfnamefont
  {J.}~\bibnamefont {Link}}, \bibinfo {author} {\bibfnamefont {D.}~\bibnamefont
  {Menze}}, \bibinfo {author} {\bibfnamefont {W.}~\bibnamefont {Neuerburg}},
  \bibinfo {author} {\bibfnamefont {M.}~\bibnamefont {Ostrick}}, \bibinfo
  {author} {\bibfnamefont {E.}~\bibnamefont {Paul}}, \bibinfo {author}
  {\bibfnamefont {I.}~\bibnamefont {Schulday}}, \bibinfo {author}
  {\bibfnamefont {W.~J.}\ \bibnamefont {Schwille}}, \bibinfo {author}
  {\bibfnamefont {H.~v.}\ \bibnamefont {Pee}}, \bibinfo {author} {\bibfnamefont
  {F.~W.}\ \bibnamefont {Wieland}}, \bibinfo {author} {\bibfnamefont
  {J.}~\bibnamefont {Wi^^c3^^9fkirchen}},\ and\ \bibinfo {author}
  {\bibfnamefont {C.}~\bibnamefont {Wu}},\ }\href
  {https://doi.org/10.1140/epja/i2003-10119-x} {\bibfield  {journal} {\bibinfo
  {journal} {Eur. Phys. J. A}\ }\textbf {\bibinfo {volume} {19}},\ \bibinfo
  {pages} {251} (\bibinfo {year} {2004})}\BibitemShut {NoStop}%
\bibitem [{\citenamefont {Sumihama}\ \emph {et~al.}(2006)\citenamefont
  {Sumihama}, \citenamefont {Ahn}, \citenamefont {Akimune}, \citenamefont
  {Asano}, \citenamefont {Bennhold}, \citenamefont {Chang}, \citenamefont
  {Corthals}, \citenamefont {Dat^^c3^^a9}, \citenamefont {Ejiri}, \citenamefont
  {Fujimura}, \citenamefont {Fujiwara}, \citenamefont {Guidal}, \citenamefont
  {Hicks}, \citenamefont {Hotta}, \citenamefont {Imai} \emph
  {et~al.}}]{sumihama__2006}%
  \BibitemOpen
  \bibfield  {author} {\bibinfo {author} {\bibfnamefont {M.}~\bibnamefont
  {Sumihama}}, \bibinfo {author} {\bibfnamefont {J.~K.}\ \bibnamefont {Ahn}},
  \bibinfo {author} {\bibfnamefont {H.}~\bibnamefont {Akimune}}, \bibinfo
  {author} {\bibfnamefont {Y.}~\bibnamefont {Asano}}, \bibinfo {author}
  {\bibfnamefont {C.}~\bibnamefont {Bennhold}}, \bibinfo {author}
  {\bibfnamefont {W.~C.}\ \bibnamefont {Chang}}, \bibinfo {author}
  {\bibfnamefont {T.}~\bibnamefont {Corthals}}, \bibinfo {author}
  {\bibfnamefont {S.}~\bibnamefont {Dat^^c3^^a9}}, \bibinfo {author}
  {\bibfnamefont {H.}~\bibnamefont {Ejiri}}, \bibinfo {author} {\bibfnamefont
  {H.}~\bibnamefont {Fujimura}}, \bibinfo {author} {\bibfnamefont
  {M.}~\bibnamefont {Fujiwara}}, \bibinfo {author} {\bibfnamefont
  {M.}~\bibnamefont {Guidal}}, \bibinfo {author} {\bibfnamefont
  {K.}~\bibnamefont {Hicks}}, \bibinfo {author} {\bibfnamefont
  {T.}~\bibnamefont {Hotta}}, \bibinfo {author} {\bibfnamefont
  {K.}~\bibnamefont {Imai}}, \emph {et~al.},\ }\href
  {https://doi.org/10.1103/PhysRevC.73.035214} {\bibfield  {journal} {\bibinfo
  {journal} {Phys. Rev. C}\ }\textbf {\bibinfo {volume} {73}},\ \bibinfo
  {pages} {035214} (\bibinfo {year} {2006})}\BibitemShut {NoStop}%
\bibitem [{\citenamefont {Lleres}\ \emph {et~al.}(2007)\citenamefont {Lleres},
  \citenamefont {Bartalini}, \citenamefont {Bellini}, \citenamefont {Bocquet},
  \citenamefont {Calvat}, \citenamefont {Capogni}, \citenamefont {Casano},
  \citenamefont {Castoldi}, \citenamefont {D'Angelo}, \citenamefont {Didelez},
  \citenamefont {Di~Salvo}, \citenamefont {Fantini}, \citenamefont {Gaulard},
  \citenamefont {Gervino}, \citenamefont {Ghio} \emph
  {et~al.}}]{lleres_polarization_2007}%
  \BibitemOpen
  \bibfield  {author} {\bibinfo {author} {\bibfnamefont {A.}~\bibnamefont
  {Lleres}}, \bibinfo {author} {\bibfnamefont {O.}~\bibnamefont {Bartalini}},
  \bibinfo {author} {\bibfnamefont {V.}~\bibnamefont {Bellini}}, \bibinfo
  {author} {\bibfnamefont {J.~P.}\ \bibnamefont {Bocquet}}, \bibinfo {author}
  {\bibfnamefont {P.}~\bibnamefont {Calvat}}, \bibinfo {author} {\bibfnamefont
  {M.}~\bibnamefont {Capogni}}, \bibinfo {author} {\bibfnamefont
  {L.}~\bibnamefont {Casano}}, \bibinfo {author} {\bibfnamefont
  {M.}~\bibnamefont {Castoldi}}, \bibinfo {author} {\bibfnamefont
  {A.}~\bibnamefont {D'Angelo}}, \bibinfo {author} {\bibfnamefont {J.~P.}\
  \bibnamefont {Didelez}}, \bibinfo {author} {\bibfnamefont {R.}~\bibnamefont
  {Di~Salvo}}, \bibinfo {author} {\bibfnamefont {A.}~\bibnamefont {Fantini}},
  \bibinfo {author} {\bibfnamefont {C.}~\bibnamefont {Gaulard}}, \bibinfo
  {author} {\bibfnamefont {G.}~\bibnamefont {Gervino}}, \bibinfo {author}
  {\bibfnamefont {F.}~\bibnamefont {Ghio}}, \emph {et~al.},\ }\href
  {https://doi.org/10.1140/epja/i2006-10167-8} {\bibfield  {journal} {\bibinfo
  {journal} {Eur. Phys. J. A}\ }\textbf {\bibinfo {volume} {31}},\ \bibinfo
  {pages} {79} (\bibinfo {year} {2007})}\BibitemShut {NoStop}%
\bibitem [{\citenamefont {Alcorn}\ \emph {et~al.}(2004)\citenamefont {Alcorn},
  \citenamefont {Anderson}, \citenamefont {Aniol}, \citenamefont {Annand},
  \citenamefont {Auerbach}, \citenamefont {Arrington}, \citenamefont {Averett},
  \citenamefont {Baker}, \citenamefont {Baylac}, \citenamefont {Beise},
  \citenamefont {Berthot}, \citenamefont {Bertin}, \citenamefont {Bertozzi},
  \citenamefont {Bimbot}, \citenamefont {Black} \emph
  {et~al.}}]{alcorn_basic_2004}%
  \BibitemOpen
  \bibfield  {author} {\bibinfo {author} {\bibfnamefont {J.}~\bibnamefont
  {Alcorn}}, \bibinfo {author} {\bibfnamefont {B.}~\bibnamefont {Anderson}},
  \bibinfo {author} {\bibfnamefont {K.}~\bibnamefont {Aniol}}, \bibinfo
  {author} {\bibfnamefont {J.}~\bibnamefont {Annand}}, \bibinfo {author}
  {\bibfnamefont {L.}~\bibnamefont {Auerbach}}, \bibinfo {author}
  {\bibfnamefont {J.}~\bibnamefont {Arrington}}, \bibinfo {author}
  {\bibfnamefont {T.}~\bibnamefont {Averett}}, \bibinfo {author} {\bibfnamefont
  {F.}~\bibnamefont {Baker}}, \bibinfo {author} {\bibfnamefont
  {M.}~\bibnamefont {Baylac}}, \bibinfo {author} {\bibfnamefont
  {E.}~\bibnamefont {Beise}}, \bibinfo {author} {\bibfnamefont
  {J.}~\bibnamefont {Berthot}}, \bibinfo {author} {\bibfnamefont
  {P.}~\bibnamefont {Bertin}}, \bibinfo {author} {\bibfnamefont
  {W.}~\bibnamefont {Bertozzi}}, \bibinfo {author} {\bibfnamefont
  {L.}~\bibnamefont {Bimbot}}, \bibinfo {author} {\bibfnamefont
  {T.}~\bibnamefont {Black}}, \emph {et~al.},\ }\href
  {https://doi.org/10.1016/j.nima.2003.11.415} {\bibfield  {journal} {\bibinfo
  {journal} {Nucl. Instrum. Methods, Phys. Res. Sect. A}\ }\textbf {\bibinfo
  {volume} {522}},\ \bibinfo {pages} {294} (\bibinfo {year}
  {2004})}\BibitemShut {NoStop}%
\bibitem [{\citenamefont {Santiesteban}\ \emph {et~al.}(2019)\citenamefont
  {Santiesteban}, \citenamefont {Alsalmi}, \citenamefont {Meekins},
  \citenamefont {Gayoso}, \citenamefont {Bane}, \citenamefont {Barcus},
  \citenamefont {Campbell}, \citenamefont {Castellanos}, \citenamefont
  {Cruz-Torres}, \citenamefont {Dai}, \citenamefont {Hague}, \citenamefont
  {Hauenstein}, \citenamefont {Higinbotham}, \citenamefont {Holt},
  \citenamefont {Kutz} \emph {et~al.}}]{santiesteban_density_2019}%
  \BibitemOpen
  \bibfield  {author} {\bibinfo {author} {\bibfnamefont {S.}~\bibnamefont
  {Santiesteban}}, \bibinfo {author} {\bibfnamefont {S.}~\bibnamefont
  {Alsalmi}}, \bibinfo {author} {\bibfnamefont {D.}~\bibnamefont {Meekins}},
  \bibinfo {author} {\bibfnamefont {C.~A.}\ \bibnamefont {Gayoso}}, \bibinfo
  {author} {\bibfnamefont {J.}~\bibnamefont {Bane}}, \bibinfo {author}
  {\bibfnamefont {S.}~\bibnamefont {Barcus}}, \bibinfo {author} {\bibfnamefont
  {J.}~\bibnamefont {Campbell}}, \bibinfo {author} {\bibfnamefont
  {J.}~\bibnamefont {Castellanos}}, \bibinfo {author} {\bibfnamefont
  {R.}~\bibnamefont {Cruz-Torres}}, \bibinfo {author} {\bibfnamefont
  {H.}~\bibnamefont {Dai}}, \bibinfo {author} {\bibfnamefont {T.}~\bibnamefont
  {Hague}}, \bibinfo {author} {\bibfnamefont {F.}~\bibnamefont {Hauenstein}},
  \bibinfo {author} {\bibfnamefont {D.}~\bibnamefont {Higinbotham}}, \bibinfo
  {author} {\bibfnamefont {R.}~\bibnamefont {Holt}}, \bibinfo {author}
  {\bibfnamefont {T.}~\bibnamefont {Kutz}}, \emph {et~al.},\ }\href
  {https://doi.org/10.1016/j.nima.2019.06.025} {\bibfield  {journal} {\bibinfo
  {journal} {Nucl. Instrum. Methods, Phys. Res. Sect. A}\ }\textbf {\bibinfo
  {volume} {940}},\ \bibinfo {pages} {351} (\bibinfo {year}
  {2019})}\BibitemShut {NoStop}%
\bibitem [{noa()}]{noauthor_httpshallcwebjlaborgwikiindexphpsimc__nodate}%
  \BibitemOpen
  \href {https://hallcweb.jlab.org/wiki/index.php/SIMC_Monte_Carlo} {\bibinfo
  {title} {https://hallcweb.jlab.org/wiki/index.php/{SIMC}\_
  {Monte}\_carlo}}\BibitemShut {NoStop}%
\bibitem [{\citenamefont {Vanderhaeghen}\ \emph {et~al.}(2000)\citenamefont
  {Vanderhaeghen}, \citenamefont {Friedrich}, \citenamefont {Lhuillier},
  \citenamefont {Marchand}, \citenamefont {Van~Hoorebeke},\ and\ \citenamefont
  {Van~de Wiele}}]{vanderhaeghen_qed_2000}%
  \BibitemOpen
  \bibfield  {author} {\bibinfo {author} {\bibfnamefont {M.}~\bibnamefont
  {Vanderhaeghen}}, \bibinfo {author} {\bibfnamefont {J.~M.}\ \bibnamefont
  {Friedrich}}, \bibinfo {author} {\bibfnamefont {D.}~\bibnamefont
  {Lhuillier}}, \bibinfo {author} {\bibfnamefont {D.}~\bibnamefont {Marchand}},
  \bibinfo {author} {\bibfnamefont {L.}~\bibnamefont {Van~Hoorebeke}},\ and\
  \bibinfo {author} {\bibfnamefont {J.}~\bibnamefont {Van~de Wiele}},\ }\href
  {https://doi.org/10.1103/PhysRevC.62.025501} {\bibfield  {journal} {\bibinfo
  {journal} {Phys. Rev. C}\ }\textbf {\bibinfo {volume} {62}},\ \bibinfo
  {pages} {025501} (\bibinfo {year} {2000})}\BibitemShut {NoStop}%
\bibitem [{\citenamefont {Tsai}(1974)}]{tsai_pair_1974}%
  \BibitemOpen
  \bibfield  {author} {\bibinfo {author} {\bibfnamefont {Y.-S.}\ \bibnamefont
  {Tsai}},\ }\href {https://doi.org/10.1103/RevModPhys.46.815} {\bibfield
  {journal} {\bibinfo  {journal} {Rev. Mod. Phys.}\ }\textbf {\bibinfo {volume}
  {46}},\ \bibinfo {pages} {815} (\bibinfo {year} {1974})}\BibitemShut
  {NoStop}%
\bibitem [{\citenamefont {Bebek}\ \emph {et~al.}(1977)\citenamefont {Bebek},
  \citenamefont {Brown}, \citenamefont {Bucksbaum}, \citenamefont {Herzlinger},
  \citenamefont {Holmes}, \citenamefont {Lichtenstein}, \citenamefont {Pipkin},
  \citenamefont {Raither},\ and\ \citenamefont
  {Sisterson}}]{bebek_electroproduction_1977}%
  \BibitemOpen
  \bibfield  {author} {\bibinfo {author} {\bibfnamefont {C.~J.}\ \bibnamefont
  {Bebek}}, \bibinfo {author} {\bibfnamefont {C.~N.}\ \bibnamefont {Brown}},
  \bibinfo {author} {\bibfnamefont {P.}~\bibnamefont {Bucksbaum}}, \bibinfo
  {author} {\bibfnamefont {M.}~\bibnamefont {Herzlinger}}, \bibinfo {author}
  {\bibfnamefont {S.~D.}\ \bibnamefont {Holmes}}, \bibinfo {author}
  {\bibfnamefont {C.~A.}\ \bibnamefont {Lichtenstein}}, \bibinfo {author}
  {\bibfnamefont {F.~M.}\ \bibnamefont {Pipkin}}, \bibinfo {author}
  {\bibfnamefont {S.~W.}\ \bibnamefont {Raither}},\ and\ \bibinfo {author}
  {\bibfnamefont {L.~K.}\ \bibnamefont {Sisterson}},\ }\href
  {https://doi.org/10.1103/PhysRevD.15.594} {\bibfield  {journal} {\bibinfo
  {journal} {Phys. Rev. D}\ }\textbf {\bibinfo {volume} {15}},\ \bibinfo
  {pages} {594} (\bibinfo {year} {1977})}\BibitemShut {NoStop}%
\bibitem [{\citenamefont {Brown}\ \emph {et~al.}(1972)\citenamefont {Brown},
  \citenamefont {Canizares}, \citenamefont {Cooper}, \citenamefont {Eisner},
  \citenamefont {Feldman}, \citenamefont {Lichtenstein}, \citenamefont {Litt},
  \citenamefont {Lockeretz}, \citenamefont {Montana}, \citenamefont {Pipkin},\
  and\ \citenamefont {Hicks}}]{brown_coincidence_1972}%
  \BibitemOpen
  \bibfield  {author} {\bibinfo {author} {\bibfnamefont {C.~N.}\ \bibnamefont
  {Brown}}, \bibinfo {author} {\bibfnamefont {C.~R.}\ \bibnamefont
  {Canizares}}, \bibinfo {author} {\bibfnamefont {W.~E.}\ \bibnamefont
  {Cooper}}, \bibinfo {author} {\bibfnamefont {A.~M.}\ \bibnamefont {Eisner}},
  \bibinfo {author} {\bibfnamefont {G.~J.}\ \bibnamefont {Feldman}}, \bibinfo
  {author} {\bibfnamefont {C.~A.}\ \bibnamefont {Lichtenstein}}, \bibinfo
  {author} {\bibfnamefont {L.}~\bibnamefont {Litt}}, \bibinfo {author}
  {\bibfnamefont {W.}~\bibnamefont {Lockeretz}}, \bibinfo {author}
  {\bibfnamefont {V.~B.}\ \bibnamefont {Montana}}, \bibinfo {author}
  {\bibfnamefont {F.~M.}\ \bibnamefont {Pipkin}},\ and\ \bibinfo {author}
  {\bibfnamefont {N.}~\bibnamefont {Hicks}},\ }\href
  {https://doi.org/10.1103/PhysRevLett.28.1086} {\bibfield  {journal} {\bibinfo
   {journal} {Phys. Rev. Lett.}\ }\textbf {\bibinfo {volume} {28}},\ \bibinfo
  {pages} {1086} (\bibinfo {year} {1972})}\BibitemShut {NoStop}%
\bibitem [{\citenamefont {Bebek}\ \emph {et~al.}(1974)\citenamefont {Bebek},
  \citenamefont {Brown}, \citenamefont {Herzlinger}, \citenamefont {Holmes},
  \citenamefont {Lichtenstein}, \citenamefont {Pipkin}, \citenamefont
  {Sisterson}, \citenamefont {Andrews}, \citenamefont {Berkelman},
  \citenamefont {Cassel}, \citenamefont {Hartill},\ and\ \citenamefont
  {Hicks}}]{bebek_electroproduction_1974}%
  \BibitemOpen
  \bibfield  {author} {\bibinfo {author} {\bibfnamefont {C.~J.}\ \bibnamefont
  {Bebek}}, \bibinfo {author} {\bibfnamefont {C.~N.}\ \bibnamefont {Brown}},
  \bibinfo {author} {\bibfnamefont {M.}~\bibnamefont {Herzlinger}}, \bibinfo
  {author} {\bibfnamefont {S.}~\bibnamefont {Holmes}}, \bibinfo {author}
  {\bibfnamefont {C.~A.}\ \bibnamefont {Lichtenstein}}, \bibinfo {author}
  {\bibfnamefont {F.~M.}\ \bibnamefont {Pipkin}}, \bibinfo {author}
  {\bibfnamefont {L.~K.}\ \bibnamefont {Sisterson}}, \bibinfo {author}
  {\bibfnamefont {D.}~\bibnamefont {Andrews}}, \bibinfo {author} {\bibfnamefont
  {K.}~\bibnamefont {Berkelman}}, \bibinfo {author} {\bibfnamefont {D.~G.}\
  \bibnamefont {Cassel}}, \bibinfo {author} {\bibfnamefont {D.~L.}\
  \bibnamefont {Hartill}},\ and\ \bibinfo {author} {\bibfnamefont
  {N.}~\bibnamefont {Hicks}},\ }\href
  {https://doi.org/10.1103/PhysRevLett.32.21} {\bibfield  {journal} {\bibinfo
  {journal} {Phys. Rev. Lett.}\ }\textbf {\bibinfo {volume} {32}},\ \bibinfo
  {pages} {21} (\bibinfo {year} {1974})}\BibitemShut {NoStop}%
\bibitem [{\citenamefont {Mohring}\ \emph {et~al.}(2003)\citenamefont
  {Mohring}, \citenamefont {Abbott}, \citenamefont {Ahmidouch}, \citenamefont
  {Amatuni}, \citenamefont {Ambrozewicz}, \citenamefont {Angelescu},
  \citenamefont {Armstrong}, \citenamefont {Arrington}, \citenamefont
  {Assamagan}, \citenamefont {Avery}, \citenamefont {Bailey}, \citenamefont
  {Beard}, \citenamefont {Beedoe}, \citenamefont {Beise}, \citenamefont
  {Breuer} \emph {et~al.}}]{mohring_separation_2003}%
  \BibitemOpen
  \bibfield  {author} {\bibinfo {author} {\bibfnamefont {R.~M.}\ \bibnamefont
  {Mohring}}, \bibinfo {author} {\bibfnamefont {D.}~\bibnamefont {Abbott}},
  \bibinfo {author} {\bibfnamefont {A.}~\bibnamefont {Ahmidouch}}, \bibinfo
  {author} {\bibfnamefont {T.~A.}\ \bibnamefont {Amatuni}}, \bibinfo {author}
  {\bibfnamefont {P.}~\bibnamefont {Ambrozewicz}}, \bibinfo {author}
  {\bibfnamefont {T.}~\bibnamefont {Angelescu}}, \bibinfo {author}
  {\bibfnamefont {C.~S.}\ \bibnamefont {Armstrong}}, \bibinfo {author}
  {\bibfnamefont {J.}~\bibnamefont {Arrington}}, \bibinfo {author}
  {\bibfnamefont {K.}~\bibnamefont {Assamagan}}, \bibinfo {author}
  {\bibfnamefont {S.}~\bibnamefont {Avery}}, \bibinfo {author} {\bibfnamefont
  {K.}~\bibnamefont {Bailey}}, \bibinfo {author} {\bibfnamefont
  {K.}~\bibnamefont {Beard}}, \bibinfo {author} {\bibfnamefont
  {S.}~\bibnamefont {Beedoe}}, \bibinfo {author} {\bibfnamefont {E.~J.}\
  \bibnamefont {Beise}}, \bibinfo {author} {\bibfnamefont {H.}~\bibnamefont
  {Breuer}}, \emph {et~al.},\ }\href
  {https://doi.org/10.1103/PhysRevC.67.055205} {\bibfield  {journal} {\bibinfo
  {journal} {Phys. Rev. C}\ }\textbf {\bibinfo {volume} {67}},\ \bibinfo
  {pages} {055205} (\bibinfo {year} {2003})}\BibitemShut {NoStop}%
\bibitem [{\citenamefont {{A1 Collaboration}}\ \emph
  {et~al.}(2012)\citenamefont {{A1 Collaboration}}, \citenamefont {Achenbach},
  \citenamefont {Ayerbe~Gayoso}, \citenamefont {Bernauer}, \citenamefont
  {Bianchin}, \citenamefont {B^^c3^^b6hm}, \citenamefont {Borodina},
  \citenamefont {Bosnar}, \citenamefont {B^^c3^^b6sz}, \citenamefont {Bozkurt},
  \citenamefont {Byd^^c5^^beovsk^^c3^^bd}, \citenamefont {Debenjak},
  \citenamefont {Distler}, \citenamefont {Esser}, \citenamefont
  {Fri^^c5^^a1^^c4^^8di^^c4^^87} \emph
  {et~al.}}]{a1_collaboration_exclusive_2012}%
  \BibitemOpen
  \bibfield  {author} {\bibinfo {author} {\bibnamefont {{A1 Collaboration}}},
  \bibinfo {author} {\bibfnamefont {P.}~\bibnamefont {Achenbach}}, \bibinfo
  {author} {\bibfnamefont {C.}~\bibnamefont {Ayerbe~Gayoso}}, \bibinfo {author}
  {\bibfnamefont {J.~C.}\ \bibnamefont {Bernauer}}, \bibinfo {author}
  {\bibfnamefont {S.}~\bibnamefont {Bianchin}}, \bibinfo {author}
  {\bibfnamefont {R.}~\bibnamefont {B^^c3^^b6hm}}, \bibinfo {author}
  {\bibfnamefont {O.}~\bibnamefont {Borodina}}, \bibinfo {author}
  {\bibfnamefont {D.}~\bibnamefont {Bosnar}}, \bibinfo {author} {\bibfnamefont
  {M.}~\bibnamefont {B^^c3^^b6sz}}, \bibinfo {author} {\bibfnamefont
  {V.}~\bibnamefont {Bozkurt}}, \bibinfo {author} {\bibfnamefont
  {P.}~\bibnamefont {Byd^^c5^^beovsk^^c3^^bd}}, \bibinfo {author}
  {\bibfnamefont {L.}~\bibnamefont {Debenjak}}, \bibinfo {author}
  {\bibfnamefont {M.~O.}\ \bibnamefont {Distler}}, \bibinfo {author}
  {\bibfnamefont {A.}~\bibnamefont {Esser}}, \bibinfo {author} {\bibfnamefont
  {I.}~\bibnamefont {Fri^^c5^^a1^^c4^^8di^^c4^^87}}, \emph {et~al.},\ }\href
  {https://doi.org/10.1140/epja/i2012-12014-9} {\bibfield  {journal} {\bibinfo
  {journal} {Eur. Phys. J. A}\ }\textbf {\bibinfo {volume} {48}},\ \bibinfo
  {pages} {14} (\bibinfo {year} {2012})}\BibitemShut {NoStop}%
\end{thebibliography}%
\end{document}